\documentclass[english,12pt,sort&compress]{iopart}
\usepackage{float}
\usepackage{iopams}
\usepackage{setstack}
\usepackage{graphicx}
\usepackage{amssymb}
\usepackage{verbatim}
\usepackage{booktabs}
\usepackage{array}
\usepackage[numbers,merge]{natbib}

\usepackage{babel}

\newcommand{\figtext}[1]{#1}

\newcommand{\fig}{Fig.\ }
\newcommand{\sect}{Sec.\ }
\newcommand{\eqn}{Eq.\ }
\newcommand{\eqns}{Eqs.\ }

\newcommand{\bigOtwo}[3]{O(#1^{#2},#1^{#3})}
\newcommand{\myL}{\Lambda}

\def\clap#1{\hbox to 0pt{\hss#1\hss}}

\begin{document}


\title[Diffusion in a logarithmic potential: scaling and selection]
{Diffusion in a logarithmic potential: scaling and
selection in the approach to equilibrium}
\author{Ori Hirschberg$^1$, David Mukamel$^1$, Gunter M. Sch{\"u}tz$^2$}

\address{$^1$ Department of Physics of Complex Systems, Weizmann
Institute of Science, Rehovot 76100, Israel}
\address{$^2$ Theoretical Soft Matter and Biophysics, Institute of
Complex Systems, Forschungszentrum J\"ulich, 52425 J\"ulich,
Germany}

\ead{\mailto{ori.hirschberg@weizmann.ac.il},
\mailto{david.mukamel@weizmann.ac.il},
\mailto{g.schuetz@fz-juelich.de}}

\begin{abstract}
The equation which describes a particle diffusing in a logarithmic
potential arises in diverse physical problems such as momentum
diffusion of atoms in optical traps, condensation processes, and
denaturation of DNA molecules. A detailed study of the approach of
such systems to equilibrium via a scaling analysis is carried out,
revealing three surprising features: (i) the solution is given by
two distinct scaling forms, corresponding to a diffusive
($x\sim\sqrt{t}$) and a subdiffusive ($x \ll \sqrt{t}$) length
scales, respectively; (ii) the scaling exponents and scaling
functions corresponding to both regimes are selected by the initial
condition; and (iii) this dependence on the initial condition
manifests a ``phase transition'' from a regime in which the scaling
solution depends on the initial condition to a regime in which it is
independent of it. The selection mechanism which is found has many
similarities to the marginal stability mechanism which has been
widely studied in the context of fronts propagating into unstable
states. The general scaling forms are presented and their practical
and theoretical applications are discussed.
\end{abstract}


\date{\today}

\section{Introduction}

A large variety of physical problems are governed by the simple
diffusion equation which describes a Brownian particle in a
logarithmic potential. Such problems range from the momentum
spreading of cold atoms in optical traps
\cite{CohenTannoudji,Zoller1996,LutzErgodicity2003,LutzTsallis2004,OpticalTsallisExperiment2006}
to the dynamics of ``bubbles'' in denaturing DNA molecules
\cite{DnaHankeMetzler2003,BarKafriMukamel2007DNA,FogedbyMetzler2007DNA,
bar2009DNAMelting,DnaKunzLivi2007,SchutzMurthy2011DNA}, and from the
relaxation of a single particle in a fluid with long-range
interactions
\cite{BouchetDauxoisRapid2005,BouchetDauxoisLong2005,CampaEtalLongRangeReview2009}
to models describing brief awakenings in the course of a night's
sleep \cite{LoEtAlSleepWake2002}. In these examples, as well as in
others which will be described below, one is interested in the
distribution of a fluctuating quantity $x$: for optically trapped
cold atoms  $x$ stands for the momentum of the atom, in the problem
of DNA denaturation $x$ is the length of a denatured unbound loop in
the double stranded molecule, and when modeling sleep dynamics, $x$
represents the wakefulness level of a sleeping individual. In the
problems we consider, the temporal evolution of the distribution
$P(x,t)$ can be approximated when $x$ is large enough by the
equation
\begin{equation}\label{eq:FokkerPlanckLargeX}
\frac{\partial P(x,t)}{\partial t} = \frac{\partial}{\partial
x}\biggl[\frac{b}{x}P(x,t)\biggr] + \frac{\partial^2P(x,t)}{\partial
x^2}.
\end{equation}
The dimensionless parameter $b$, which plays a central role in the
solution of the equation, has a different physical meaning in each
problem. In physical systems of the type which we consider, \eqn
(\ref{eq:FokkerPlanckLargeX}) has corrections at small values of
$x$, and, in particular, the divergence at small $x$ is not present.

To have a concrete physical picture in mind, we will mostly
concentrate on a Brownian particle diffusing under the influence of
a one-dimensional external potential which is logarithmic for large
$x$:
\begin{equation}\label{eq:LogPotential}
\frac{V(|x|\gg 1)}{k_B T}  \sim b \log (|x|),
\end{equation}
where $T$ is the temperature, $k_B$ is Boltzmann's constant and $b$
measures the strength of the potential. The corresponding
Fokker-Planck equation has the form of a continuity equation for the
probability distribution
\begin{equation}\label{eq:FokkerPlanck}
\frac{\partial P(x,t)}{\partial t} = - \frac{\partial
J(x,t)}{\partial x}, \quad \mbox{with}\quad J(x,t) = -V'(x)P(x,t) -
\frac{\partial P(x,t)}{\partial x},
\end{equation}
where $J(x,t)$ denotes the probability current. To simplify
notation, here and throughout the paper time is measured in units in
which the diffusion coefficient is equal to 1, and the potential is
measured in units of temperature, i.e., $k_B T = 1$.  For large
values of $|x|$ this Fokker-Planck equation reduces to
(\ref{eq:FokkerPlanckLargeX}). For some applications it is natural
to restrict the variable $x$ to be non-negative, in which case the
equation should be supplemented by a boundary condition at $x=0$. We
therefore discuss both the case of restricted $x$ with different
boundary conditions, and the case where $x$ is unbounded (which
requires no additional boundary conditions).

Our goal is to study the long-time behavior of the solutions to this
diffusion problem. In a recent paper \cite{LogDiffPRE2011} we have
reported that the solutions of equation (\ref{eq:FokkerPlanck})
relax to equilibrium via a universal scaling form which depends on
the potential only through its asymptotic form
(\ref{eq:LogPotential}), and also depends on the initial condition.
In the present paper we elaborate on the analysis of
\cite{LogDiffPRE2011}, present in more detail the derivation of this
result, and discuss its implications and applications.

The scaling form which we find exhibits several features which are
not typically found in scaling solutions. (i) For any large finite
time, the overall scaling form is comprised of two scaling
functions, one corresponding to small values of $x$ ($x\ll
\sqrt{t}$) and another corresponding to large values of $x$ ($x
\sim\sqrt{t}$). Together, these two scaling functions give the
distribution $P(x,t)$ for all values of $x$, including the
microscopic scale where the potential deviates from a logarithm. The
small-$x$ details of the potential $V(x)$ enter only through the
steady-state distribution which multiplies this scaling form. (ii)
The equation admits not only one, but a family of such overall
scaling forms, each characterized by different scaling exponents.
The scaling form which describes the observed relaxation to
equilibrium depends on the initial condition via a selection
mechanism akin to the marginal stability selection mechanism
encountered in, e.g., fronts propagating into unstable states. (iii)
As in problems of propagating fronts, by continuously changing the
tail of the initial condition, the selected scaling exponent
exhibits a ``phase transition'' from a smoothly varying to a fixed
value.

Features (ii) and (iii) of the scaling solution provide an
interesting connection between this diffusion problem and the
well-known problem of fronts propagating into unstable states. Many
systems of the latter type admit a family of traveling-wave
solutions with different propagation velocities, and the mechanism
by which the eventual velocity and waveform are selected has been
widely studied \cite{VanSaarloosReview2003}. As described below, the
selection mechanism which we find for the diffusion problem is
similar in many of its details to that corresponding to propagating
fronts. The two problems differ however in some basic aspects:
unlike the homogeneous nonlinear propagating front problems,
equation (\ref{eq:FokkerPlanck}) is \emph{linear} yet
\emph{inhomogeneous in space}. Although the inhomogeneity of our
problem restricts the utility of mathematical methods used to
analyze the selection of front velocities, most notably Fourier
analysis, its linearity makes it exactly solvable and facilitates
the demonstration of the selection mechanism. The similarities
between the problems, suggests that a common mathematical
description of their solutions might exist.

As mentioned above, at late times the entire distribution is given
by a scaling form (feature (i) above). At late times $t \gg 1$, two
different length scales emerge: a large-$x$ length scale of $x \sim
t^{1/2}$ and and a small-$x$ length scale of $x\sim t^\gamma$ with a
$b$-dependent exponent $\gamma<1/2$. Hereafter we refer to these two
length scales as ``the large-$x$'' and ``the small-$x$'' regions,
respectively. The exponent $\gamma$ depends on the boundary
conditions, and in particular, for a reflecting boundary condition
at the origin, $\gamma = 1/(b+1)$. The solution in each of the
length scales is given by a different scaling function, with a
smooth interpolation between the two functions. Moreover, to leading
order in $t$, these two scaling functions yield the solution at
\emph{any} point $x$ (see \fig \ref{fig:SchematicScaling}). Both
scaling functions are selected by the initial condition: the
large-$x$ scaling function determines the one in the small-$x$
region. In the language of traveling waves, this corresponds to a
system with two fronts propagating with different velocities,
whereby the selected scaling solution of the ``faster front''
dictates that of the ``slower'' one.

For a wide class of initial conditions, which include compactly
supported (or ``steep'') initial distributions, the large-$x$
scaling solution of \eqns
(\ref{eq:LogPotential})--(\ref{eq:FokkerPlanck}) has recently been
found in \cite{GodrecheLuck2001ZetaUrn}, \cite{ProbeParticles} and
\cite{BarkaiKessler}. There, the dependence of the solution on the
initial condition and the behavior at small $x$ have not been
considered. In fact, in many physical circumstances the initial
distribution is not steep. In other situations, the small-$x$
behavior rather than the large-$x$ one determines the physical
observables of interest. In these two respects, beyond the relevance
of our work to the general theory of selection problems and scaling
solutions, it also presents a comprehensive analysis of the scaling
solution of (\ref{eq:LogPotential})--(\ref{eq:FokkerPlanck}) which
provides useful results for many concrete systems. To demonstrate
the applicability of our results, we consider at the end of this
paper three physical examples: (a) cold atoms in an optical lattice
undergoing a rapid ``quench'' from one steady state to another. Here
we discuss initial conditions with a fat tail. (b) Nonequilibrium
driven models exhibiting real-space condensation. Here we show that
current correlations in these systems may be evaluated by
considering initial conditions with specific algebraic decay at the
tails. (c) The dynamics of loops in DNA molecules undergoing
denaturation. Here the effect of an absorbing boundary condition is
probed.

The paper is organized as follows. In \sect \ref{sec:Scaling} we
present the scaling solution and its selection mechanism, and
heuristically derive its form. The results of this section are
backed up by an exact solution of the Fokker-Planck equation
(\ref{eq:FokkerPlanckLargeX}) which appears in
\ref{sec:calculation}. In \sect \ref{sec:CommentsOnScaling}, the
scaling solution is discussed in the broader contexts of the general
theory of scaling solutions (\sect
\ref{sec:IncompleteSelfSimilarity}), of selection in problems of
propagating fronts (\sect \ref{sec:SelectionFronts}), and of the
results of previous work on \eqn (\ref{eq:FokkerPlanckLargeX})
(\sect \ref{sec:PreviousWork}). The discussion in Sections
\ref{sec:Scaling} and \ref{sec:CommentsOnScaling} is focused on
systems whose boundary conditions conserve probability. In \sect
\ref{sec:NonConservingBC} we generalize the results of the previous
sections to the case in which probability is not conserved at the
boundary. In particular, we show that the large-$|x|$ scaling form
is not affected by the boundary conditions. Applications of our
results to concrete physical systems are discussed in \sect
\ref{sec:applications}, in which we also present a general review of
some of the problems which are described by \eqns
(\ref{eq:LogPotential})--(\ref{eq:FokkerPlanck}), for which our
results may be relevant. Finally, \sect \ref{sec:conclusion}
contains a summary of our results and some concluding remarks.

\section{The scaling solution and its
universal character}\label{sec:Scaling}

In this section we present the scaling solution of \eqn
(\ref{eq:LogPotential})--(\ref{eq:FokkerPlanck}). We begin in \sect
\ref{sec:GeneralScaling} with a general discussion of the problem of
diffusion in a logarithmic potential, and present its scaling
solution. In the following subsections this results is derived
heuristically, while the exact derivation of this result, which is
somewhat technical, is found in \ref{sec:calculation}. First, in
\sect \ref{sec:ScalingLargeX} we demonstrate that in the large-$x$
regime of $x\sim \sqrt{t}$, \eqn (\ref{eq:FokkerPlanckLargeX})
admits a one-parameter family of scaling solutions. In \sect
\ref{sec:selection}, we present the selection criterion which
explains how the initial conditions determine which member of this
family is eventually observed. In \sect \ref{sec:ScalingSmallX} we
derive the scaling form for the small-$x$ regime. The derivation of
\sect \ref{sec:ScalingLargeX} rests on the assumption that when
$x\sim \sqrt{t}\gg 1$, \eqn (\ref{eq:FokkerPlanck}) can be well
approximated by \eqn (\ref{eq:FokkerPlanckLargeX}). In \sect
\ref{sec:universality} we justify this assumption by showing that
our scaling solution is universal, i.e., it depends only on the
large-$|x|$ tails of the potential (\ref{eq:LogPotential}).

\subsection{General discussion of the problem}\label{sec:GeneralScaling}

We begin by writing the Fokker-Planck equation
(\ref{eq:LogPotential})--(\ref{eq:FokkerPlanck}) in more concrete
terms. We consider a potential has of form
\begin{equation}\label{eq:LogPotentialFull}
V(x) = b \log (|x|) + U(x),
\end{equation}
where the correction $U(x)$ is negligible for large $|x|$ and it
ensures that $V(x)$ does not diverge at the origin. For
concreteness, we assume that for large $x$
\begin{equation}\label{eq:CorrectionToScaling}
U(x\gg 1)  = O(|x|^{-\sigma})\quad \mbox{with}\quad \sigma>0.
\end{equation}
Regularizing the potential at small $x$ is needed since for $b>1$,
the case on which we focus below, a logarithmic divergence of the
potential at the origin makes $x=0$ an absorbing state and any
normalized initial condition tends to a $\delta$-function
distribution around it
\cite{FirstPassageBesselProcess2011,FogedbyFiniteTimeSingularities2002}.
This suggests that in systems with $b>1$, physical corrections to
the logarithmic potential near the origin cannot be neglected when
analyzing the long-time behavior.

With this notation, the Fokker-Planck equation
(\ref{eq:FokkerPlanck}) is
\begin{equation}\label{eq:FokkerPlanckDetailed}
\frac{\partial P(x,t)}{\partial t} = \frac{\partial}{\partial
x}\biggl[\frac{b}{x}\bigl(1+h(x)\bigr)P(x,t)\biggr] +
\frac{\partial^2P(x,t)}{\partial x^2}
\end{equation}
where
\begin{equation}\label{eq:ForceCorrection}
h(x) \equiv \frac{xU'(x)}{b} = O(|x|^{-\sigma}).
\end{equation}
We begin in this section by considering only the boundary condition
at $x=0$ where the probability flux at the origin is zero, i.e.,
$J(0) = 0$. This corresponds to diffusion on the entire real line
with an even initial condition, or diffusion on the positive half
line with a reflecting boundary condition at the origin. The general
case, and the effect of other boundary conditions will be examined
in \sect \ref{sec:NonConservingBC}.

The stationary solution of the diffusion equation
(\ref{eq:FokkerPlanck}) in this potential has the form of a
Boltzmann distribution
\begin{equation}\label{eq:EqDist}
P^*(x)  = \frac{1}{Z}e^{- V(x)} \sim \frac{1}{Z}x^{-b},
\end{equation}
where $Z$ is a normalization constant given by $Z = \int e^{-
V(x)}dx$. For $b>1$, $Z$ is finite and the system tends towards this
unique equilibrium distribution regardless of the initial condition
(we assume that the potential does not contain infinite energy
barriers and the system is ergodic). However, for $b\leq 1$, the
equilibrium distribution cannot be normalized. In this case, any
normalized initial condition tends to zero. Thus, potentials with
logarithmic tails are a marginal case for the diffusion equation.
Any potential which increases at large $x$ faster than a logarithm
``traps'' the particle and the probability distribution reaches a
steady state at long times. On the other hand, potentials which
increase with $x$ slower than logarithmically are non-trapping and
the probability distribution eventually spreads out to infinity. In
the marginal case where the potential is logarithmic at large $x$,
the particle is trapped at low temperatures and becomes delocalized
at high temperatures, as the dimensionless parameter $b$ changes
from $b>1$ to $b\leq 1$.

The aim of this paper is to describe how $P(x,t)$ relaxes towards
the eventual Boltzmann distribution. We therefore concentrate on the
normalizable case $b>1$. Recently,
\cite{GodrecheLuck2001ZetaUrn,ProbeParticles,BarkaiKessler} have
shown that this relaxation is given by a useful and compact scaling
form. The scaling form which was found, however, describes the
solution of \eqns (\ref{eq:LogPotential})--(\ref{eq:FokkerPlanck})
only for a specific (albeit large) class of initial conditions, and
it represents correctly only the large-$|x|$ behavior of the actual
scaling solution. The main result of the present work is the
surprising fact that the long-time scaling form of the solution
depends on the initial condition in a non-trivial fashion. Moreover,
the entire solution can be described by a scaling form, where the
non-universal features are contained in $P^*(x)$. We now present
these results, and then derive them.

As we are interested in the relaxation towards the equilibrium
distribution, it is convenient to study the deviation of $P(x,t)$
from $P^*(x)$. To this end we define a function $G(x,t)$ via
\begin{equation}\label{eq:GDefinition}
\fl \qquad P(x,t) = P^*(x)\bigl[1+G(x,t)\bigr], \quad\mbox{or
equivalently}\quad G(x,t) = \frac{P(x,t)-P^*(x)}{P^*(x)}.
\end{equation}
We seek scaling solutions of $P(x,t)-P^*(x)$, or equivalently of
$G(x,t)$, rather than of $P(x,t)$. Note that since the Fokker-Planck
equation is linear and it is satisfied by $P^*(x)$, the distribution
$P(x,t)$ and the deviation from equilibrium $P(x,t)-P^*(x) =
P^*(x)G(x,t)$ satisfy the same equation. However, while $\int
P(x,t)dx = 1$, here $\int P^*(x)G(x,t)dx = 0$. Looking for a scaling
form for solutions with zero normalization is our primary extension
of the calculations of
\cite{GodrecheLuck2001ZetaUrn,ProbeParticles,BarkaiKessler} which
enables us to find \emph{all} scaling solutions to the problem.

\subsection{The scaling solution}\label{sec:results}

The Fokker-Planck equation (\ref{eq:FokkerPlanckLargeX}) can be
solved exactly by standard methods \cite{Risken}. By a
transformation of variables, it can be mapped to a Schr\"odinger
equation in imaginary time which describes a quantum particle moving
in an inverse square potential. Analysis of this quantum problem
yields the exact solution of the equation for arbitrary initial
conditions represented as a series of Bessel functions. Asymptotic
analysis of these Bessel functions allows one to identify the
scaling form which characterizes the approach to equilibrium at late
times. Although this calculation is straightforward, it is rather
technical and lengthy. We therefore delay its presentation to
\ref{sec:calculation}. Here, we present the results of this
calculation, and in the rest of this section we heuristically
motivate these results.

According to the exact calculation of \ref{sec:calculation}, the
long-time behavior of the solution of the Fokker-Planck equation
(\ref{eq:FokkerPlanck}) with any normalizable initial condition and
for a reflecting boundary condition at the origin, is given, to
leading order in $t$, by
\begin{equation}\label{eq:FinalScalingAnsatz}
P(x,t) \approx P^*(x) + C P^*(x)\cdot \left\{
\begin{array}{ll} g_{\beta}\bigl(\frac{|x|}{t^{1/(b+1)}}\bigr)
t^{-\delta}  & \mbox{ for } |x|\leq x_1(t) \\ \\
f_{\beta} \bigl(\frac{|x|}{t^{1/2}}\bigr)t^{-\beta} & \mbox{ for }
|x|\geq x_1(t)
\end{array} \right.,
\end{equation}
where $x_1(t)$ can be chosen to have any value which satisfies
\begin{equation}\label{eq:FrontBoundaryRange}
t^{1/(b+1)} \ll x_1(t) \ll t^{1/2},
\end{equation}
and the scaling functions are given by
\begin{eqnarray}\label{eq:FinalSmallXScalingFunction}
g_\beta(z) &= - {\textstyle \frac{4(b+1)}{r Z(2\beta+b-1)}} +
z^{b+1},
\\
f_\beta(u) &= u^{b+1}\,_1\!F_1\Bigl({\textstyle
\frac{1+b+2\beta}{2}};{\textstyle \frac{b+3}{2}};{\textstyle
-\frac{u^2}{4}}\Bigr). \label{eq:FinalScalingFunction}
\end{eqnarray}
The values of the scaling exponents $\beta$ and $\delta$ and of the
constant $C$ will be discussed shortly. The constant $r$ depends on
the domain on which the diffusion is defined: $r=2$ for diffusion on
the positive half-line (with a wall at the origin), while $r=1$ for
symmetric diffusion on the entire real line. Here, $_1 F_1$ is the
confluent hypergeometric function, whose known properties yield the
asymptotic form \cite{AbramowitzStegun}
\begin{equation}\label{eq:AsymptoticF}
f_\beta(u) \sim \left\{
\begin{array}{ll} u^{b+1}\phantom{e^{-\frac{u^2}{4}}}
& \mbox{ for } u \ll 1 \\
D u^{-2\beta}\phantom{e^{-\frac{u^2}{4}}} & \mbox{ for } u \gg 1,
\beta < 1 \\
u^{b+1}e^{-\frac{u^2}{4}} & \mbox{ for } u \gg 1, \beta = 1
\end{array} \right.,
\end{equation}
where $D = \frac{2^{1+b+2\beta}\Gamma[(b+3)/2]}{\Gamma(\beta-1)}$.

\begin{figure}
  \center
  \includegraphics[width = 0.6\textwidth]{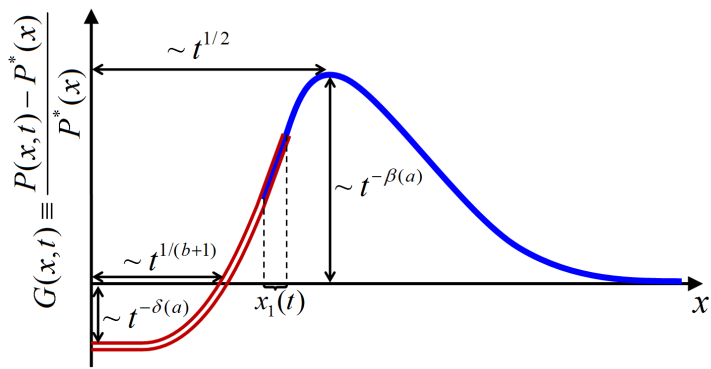}
  \caption{ \label{fig:SchematicScaling}\figtext{A schematic
  representation of the solution $P(x,t)$ (\eqn (\ref{eq:FinalScalingAnsatz}))
  at a given late time $t \gg 1$ (not drawn to scale).
  The red double line represents the
  small-$x$ scaling form $g_\beta(x/t^{1/(b+1)})t^{-\delta}$ while the blue
  solid line represents the large-$x$ scaling form
  $f_\beta(x/t^{1/2})t^{-\beta}$. The interval on which the two solutions
  overlap (\eqn (\ref{eq:FrontBoundaryRange})) is labeled $x_1(t)$.
  }}
\end{figure}

The solution (\ref{eq:FinalScalingAnsatz}) is presented
schematically in \fig \ref{fig:SchematicScaling}. It is made up of
two different scaling forms with different dynamical exponents:
$x\sim t^{1/(b+1)}$ and $x\sim t^{1/2}$. At small values of $|x|$,
the solution is flat up to $x \sim t^{1/(b+1)}$, with a value that
approaches zero as $t^{-\delta}$. At large values of $|x|$, the
solution exhibits a peak at $x\sim \sqrt{t}$, whose height shrinks
as $t^{-\beta}$ (this schematic form is modified for negative
$\beta$ since in this case $f_\beta(u)$ diverges for large $u$, see
(\ref{eq:AsymptoticF})). As \eqns
(\ref{eq:FinalSmallXScalingFunction})--(\ref{eq:AsymptoticF})
indicate, the two scaling functions are, to leading order in $t$,
identical for any $x_1(t)$ in the range
(\ref{eq:FrontBoundaryRange}), explaining why the crossover point
between the two regimes can be chosen anywhere in this range.

According to the calculation of \ref{sec:calculation}, the values of
$\beta$, $\delta$ and $C$ depend of the initial conditions. We
consider initial conditions $G_0(x) \equiv G(x,0)$ which for large
$|x|$ have an asymptotic form
\begin{equation}\label{eq:InitialAsymptotics}
G_0(|x| \gg 1) \sim A |x|^{-a}.
\end{equation}
Here, $a>1-b$ must hold for $P(x,0)$ to be normalizable. Note that
$a$ may be negative. If $G_0(x)$ decays faster than a power law, we
formally take $a=\infty$. A few examples of different initial
conditions and the corresponding values of $a$ and $A$ are given in
Table \ref{tab:InitialConditions}. In \sect
\ref{sec:NonConservingBC} we briefly discuss cases in which $G_0(x)$
is not asymptotically symmetric, i.e., when the tails of $G_0(x)$ at
$x\to\pm\infty$ decay at different rates.

\begin{table}[t!]
  \begin{center}
  \begin{tabular}{>{$}c<{\quad{}$} >{$}r<{$} @{$\;\sim\;{}$} >{$}l<{\quad{}$} >{$}c<{$} >{$}c<{$}}
    \toprule
    P(x,0) & \multicolumn{2}{>{$}c<{$}}{G_0(x)} & \; a\;{} & \;A\;{}\\
    \midrule
    \delta(x-x_0) & \frac{\delta(x-x_0)}{P^*(x)}-1 & -1 & 0 & -1 \\ \addlinespace
    C e^{-|x|/x_0} & \frac{C e^{-|x|/x_0}}{P^*(x)}-1 & -1  & 0 & -1 \\ \addlinespace
    C|x|^{-(b+1)} + \ell(x) & \frac{C|x|^{-(b+1)} + \ell(x)}{P^*(x)}-1 & -1 & 0 & -1 \\\addlinespace
    C|x|^{-(b-1)} + \ell(x) & \frac{C|x|^{-(b-1)} + \ell(x)}{P^*(x)}-1 & C Z|x|^{+1} & -1 & C Z \\\addlinespace
    P^*(x) + C|x|^{-(b+1)} + \ell(x) & \frac{C|x|^{-(b+1)} + \ell(x)}{P^*(x)}\phantom{{}-1} & C Z|x|^{-1} & 1 & C Z \\\addlinespace
    P^*\bigl(|x| + x_0\bigr) + \ell(x) & \frac{P^*(|x|+x_0) + \ell(x)}{P^*(x)} - 1 &  -b x_0|x|^{-1} & 1 & -b x_0\\\addlinespace
    P^*(x)\bigl[1 + e^{-|x|/x_0}\bigr] + \ell(x)& e^{-|x|/x_0} + \frac{\ell(x)}{P^*(x)} & e^{-|x|/x_0} & \infty & \\\addlinespace
    C P^*(x)\bigl[1 + e^{-|x|/x_0}\bigr] & C-1+Ce^{-|x|/x_0} & C - 1 & 0 & C - 1 \\
    \bottomrule
    \end{tabular}
  \caption{\figtext A variety of initial conditions $P(x,0)$ and the
  corresponding values of $a$ and $A$ according to \eqn
  (\ref{eq:InitialAsymptotics}). $G_0(x)\equiv G(x,0)$ is defined by \eqn
  (\ref{eq:GDefinition}), and its leading asymptotic behavior
  for $|x| \gg 1$ is presented. $\ell(x)$ and
  $C$ are a compactly-supported function and a constant whose values
  change from line to line. They are added to ensure the
  normalization $\int P(x,0)dx = 1$. The equilibrium distribution
  $P^*(x)$ is given in \eqn (\ref{eq:EqDist}).
  }
  \label{tab:InitialConditions}
  \end{center}
\end{table}

For this large class of initial conditions, the scaling exponents
are given by
\begin{equation}\label{eq:FinalBeta}
\beta = \beta(a) = \left\{
\begin{array}{ll} \frac{a}{2} & \mbox{if } a<2 \\
1 & \mbox{if } a>2
\end{array} \right.
\end{equation}
and
\begin{equation}\label{eq:FinalDelta}
\delta = \delta(a) = \beta(a) + \frac{b-1}{2}
\end{equation}
For $a<2$, the constant $C$ is
\begin{equation}\label{eq:FinalConstant}
C =
\frac{\Gamma(1-\frac{a}{2})}{2^{b+a+1}\Gamma(\frac{3+b}{2})}\cdot A,
\end{equation}
while for $a>2$, $C$ depends on the full forms of the initial
condition and the potential. For $a=2$ there are logarithmic
corrections to \eqn(\ref{eq:FinalScalingAnsatz}), which are
presented in \eqns (\ref{eq:ExactFinalScalingSmallX}),
(\ref{eq:ExactFinalConstant}) and (\ref{eq:ExactFinalScaling}) of
\ref{sec:calculation}.

We make two comments about the solution
(\ref{eq:FinalScalingAnsatz}). First, we would like to emphasize the
non-trivial fashion in which the solution depends on the initial
condition. The scaling functions $f_\beta$ and $g_\beta$ and the
scaling exponent $\delta$ are determined by the value of $\beta$.
The $\beta$ exponent exhibits a ``phase transition'' at $a=2$,
between a regime ($a<2$) in which $\beta$ depends on the value of
$a$ and a regime ($a>2$) in which it does not. As discussed in \sect
\ref{sec:selection} below, it is this threshold phenomenon which
ties our scaling solution with the problem of velocity selection of
propagating fronts.

The second comment is that this scaling solution is universal, in
two ways: it is independent of the small-$x$ details of the
potential, i.e., of $U(x)$ of \eqn (\ref{eq:LogPotentialFull}). It
is also independent of the small-$x$ details of the initial
condition. When we say below that a particular result is universal,
we use the term in both these meanings. To be more precise, the
universal function is $\frac{P(x,t) - P^*(x)}{P^*(x)}$; $P(x,t)$
itself depends on $U(x)$ for small values of $x$, but only through
the simple Boltzmann distribution (\ref{eq:EqDist}). It is
interesting to note that when $a>2$, where the solution does not
depend on the initial condition, the constant $C$ is non-universal,
while in the case of $a \leq 2$, where the initial condition does
affect the scaling form, $C$ is universal.

In the remainder of this section, we motivate these results in order
to gain an understanding of their origin. To do so, we derive these
results in a heuristic fashion which, although not rigorous, is more
transparent than the calculation of \ref{sec:calculation}. In this
heuristic derivation, we do not presume the results of
\ref{sec:calculation}, but for a single fact: in order to establish
the selection mechanism which leads to \eqn (\ref{eq:FinalBeta}) (in
\sect \ref{sec:selection}), we rely on the fact that localized
initial conditions $G_0(x)$ (which correspond to $a=\infty$) evolve
into scaling solutions of the form (\ref{eq:FinalScalingAnsatz})
with $\beta = 1$. In other words, using the scaling solution for
\emph{localized} initial conditions, we are able to find the scaling
solution for \emph{all} initial conditions.

\subsection{Scaling solution for $|x|\sim\sqrt{t}$ (``large $x$'')}
\label{sec:ScalingLargeX}

As we are seeking a scaling form for $G(x,t)$ (\eqn
(\ref{eq:GDefinition})) rather than for $P(x,t)$, we start by
writing down the equation governing the evolution of $G$.
Substituting (\ref{eq:GDefinition}) in the Fokker-Planck equation
(\ref{eq:FokkerPlanckDetailed}) we find
\begin{equation}\label{eq:FokkerPlanckForG}
\frac{\partial G}{\partial t} = -V'(x) \frac{\partial G}{\partial x}
+ \frac{\partial^2 G}{\partial x^2} = -\frac{b}{x}\bigl(1+h(x)\bigr)
\frac{\partial G}{\partial x} + \frac{\partial^2 G}{\partial x^2}.
\end{equation}
Here we have used \eqn (\ref{eq:EqDist}) to deduce that
$\frac{\partial P^*(x)}{\partial x} =  -V'(x) P^*(x)$. Our heuristic
derivation of the scaling solution (\ref{eq:FinalScalingAnsatz})
proceeds by dropping the $h(x)$ term in this equation, which is
negligible for large values of $x$. This leads to
\begin{equation}\label{eq:FokkerPlanckForGLargeX}
\frac{\partial G(x,t)}{\partial t} = -\frac{b}{x}\frac{\partial
G(x,t)}{\partial x} + \frac{\partial^2G(x,t)}{\partial x^2},
\end{equation}
which is equivalent to \eqn (\ref{eq:FokkerPlanckLargeX}). Dropping
$h(x)$ is justified below, in \sect \ref{sec:universality}, where we
establish the universality of the results which we now derive, i.e.,
their independence of the form of $h(x)$.

The goal of the present subsection is to show that \eqn
(\ref{eq:FokkerPlanckForGLargeX}) admits a family of scaling
solutions. We start by looking for scaling solutions of the form
\begin{equation}\label{eq:ScalingAnsatz}
G(x,t) = t^{-\beta} {f}\Bigl(\frac{|x|}{\sqrt{t}}\Bigr)
\end{equation}
where the scaling exponent $\beta$ and the function $f(u)$ are to be
determined. This corresponds to the ansatz
\begin{equation}\label{eq:ScalingAnsatzForP}
P(x,t) = P^*(x) + P^*(x) t^{-\beta}
{f}\Bigl(\frac{|x|}{\sqrt{t}}\Bigr) \sim |x|^{-b} \Bigl[1+t^{-\beta}
f\Bigl(\frac{|x|}{\sqrt{t}}\Bigr)\Bigr]
\end{equation}
for the probability distribution. Substituting
(\ref{eq:ScalingAnsatz}) in the Fokker-Planck equation
(\ref{eq:FokkerPlanckForGLargeX}) yields a family of ordinary
differential equations for $f(u)$,
\begin{equation}\label{eq:ScalingODE}
f''+\Bigl(\frac{u}{2}-\frac{b}{u}\Bigr)f' + \beta f = 0
\end{equation}
with $\beta$ a free parameter. For every value of $\beta$ this
equation has a solution
\begin{eqnarray}\label{eq:ODESolution}
f(u) = C_1
u^{b+1}\,_1\!F_1\left(\frac{1+b+2\beta}{2};\frac{b+3}{2};-\frac{u^2}{4}\right)
+ C_2\,_1\!F_1\left(\beta;\frac{1-b}{2};-\frac{u^2}{4}\right),
\nonumber \\
\end{eqnarray}
where $_1 F_1$ is the confluent hypergeometric function
\cite{AbramowitzStegun}, and $C_1$ and $C_2$ are integration
constants. The three unknown constants $\beta$, $C_1$ and $C_2$
should in principle be determined by the two boundary conditions at
$u=0$ and $u=\infty$ and by the initial condition.

The study of the small-$x$ scaling solution in \sect
\ref{sec:ScalingSmallX} below shows that the proper boundary
condition to consider at $u=0$ is
\begin{equation}\label{eq:ODEBoundaryCondition}
f(u \ll 1) \sim u^{b+1}.
\end{equation}
Using the asymptotics of the hypergeometric function
\cite{AbramowitzStegun}
\begin{equation}\label{eq:1F1Asymptotic}
\fl \qquad _1 F_1\bigl(r;s;{\textstyle -\frac{u^2}{4}}\bigr) =
\left\{
\begin{array}{ll} 1 + O(u^2)
& \mbox{ for } u \ll 1 \\
\frac{4^{r}\Gamma(s)+O(u^{-2})}{\Gamma(s-r)}\cdot
u^{-2r}\phantom{e^{-\frac{u^2}{4}}}
 & \mbox{ for } u \gg 1,\,
r-s \neq 0,1,2,\ldots  \\
\frac{\Gamma(s)+O(u^{-2})}{(-4)^{r-s}\Gamma(r)}\cdot
u^{2(r-s)}e^{-\frac{u^2}{4}} & \mbox{ for } u \gg 1,\, r-s =
0,1,2,\ldots
\end{array} \right.,
\end{equation}
we see that the boundary condition (\ref{eq:ODEBoundaryCondition})
implies that $C_2=0$, and therefore $f(u) = C f_\beta(u)$, where
$f_\beta$ is given in (\ref{eq:FinalScalingFunction}), and we have
defined $C \equiv C_1$.

Without another condition which may set the values of the two
remaining constants, $C$ and, more importantly, $\beta$, we are
still left with a family of scaling solutions. Note that the
conservation of probability cannot be used to determine these
constants, since the scaling ansatz (\ref{eq:ScalingAnsatz}) does
not hold for small enough values of $x$. Similarly, the known
stationary distribution (\ref{eq:EqDist}) does not provide a
boundary condition as all solutions relax to it (as can be seen
using (\ref{eq:1F1Asymptotic})). We therefore arrive at the uncommon
(although not unique, see \cite{BarenblattBook}) situation in which
the scaling exponent $\beta$ and the scaling function $f$ are
determined by the initial condition. This situation confronts us
with a problem of selection: which of the family of scaling
solutions is selected by the initial condition of the physical
system under consideration? We turn to this question in the next
subsection.

\subsection{Stability and the selection of the scaling
solution}\label{sec:selection}

In this section we elucidate the selection mechanism which leads to
\eqn (\ref{eq:FinalBeta}). Since probability is locally conserved by
the diffusion equation (\ref{eq:FokkerPlanck}), it is reasonable to
expect that the relaxation towards equilibrium propagates as a
diffusive ``front'' from the origin towards the tails. If this is
so, then at any given time, the tails of $G(x,t)$ do not yet
``feel'' this front, and they should therefore be given by the
initial distribution. By matching the tails of
\eqn(\ref{eq:ODESolution}) with initial conditions of the form
(\ref{eq:InitialAsymptotics}), the asymptotics
(\ref{eq:1F1Asymptotic}) of the hypergeometric function suggest that
$\beta(a) = a/2$ and $C$ is given by (\ref{eq:FinalConstant}) when
$a \neq 2,4,6,\ldots$.

According to the exact calculation of \ref{sec:calculation}, the
naive argument of the previous paragraph is correct only for $a<2$.
To understand why the argument fails when $a>2$, we turn to a
stability analysis of the scaling solutions and show that those with
$\beta > 1$ are unstable to localized perturbations. To this end, we
make use of the following result which is derived in
\ref{sec:calculation}: localized initial conditions $G(x,0)$, such
as compactly supported ones, evolve at long times to
\begin{equation}\label{eq:ScalingFunctionBeta1}
G(x,t) \sim t^{-1}f_1\Bigl(\frac{x}{\sqrt{t}}\Bigr)
\end{equation}
where $f_1(u)$ is given in (\ref{eq:FinalScalingFunction}). This
result can heuristically be understood as follows: if the initial
condition is compactly supported, then it is plausible that the
selected solution will be the one whose decay at the tails is
steepest. The asymptotics (\ref{eq:1F1Asymptotic}) of the scaling
solutions show that this is the $\beta = 1$ scaling function. We
note that using the identity
\begin{equation}\label{eq:HypergeomExpIdentity}
\,_1\!F_1 (A ;A ;z) = e^z,
\end{equation}
one can simplify the expression for the scaling function to $f_1(u)
= u^{b+1} e^{-{u^2}/{4}}$.

Let us consider a distribution which at some time $t$ is close to a
scaling solution of the form (\ref{eq:ScalingAnsatzForP}). The
distribution cannot be \emph{exactly} equal to this scaling solution
in any physical situation: at small enough $x$'s, where the
potential deviates from a logarithm, the scaling form breaks down.
At best, the exact solution is equal to the scaling solution plus a
small localized disturbance $\delta P$, i.e.,
\begin{equation}\label{eq:LocalDisturbance}
P(x,t) - P^*(x) = C P^*(x) t^{-\beta}
f_\beta\Bigl(\frac{x}{\sqrt{t}}\Bigr) + \delta P(x,t)
\end{equation}

Examining \eqn (\ref{eq:LocalDisturbance}) we see that the late-time
behavior of the solution will be close to the scaling solution
$f_\beta$ if the disturbance $\delta P$ is negligible compared to
it. In other words, at late times we can only see scaling solutions
which are \emph{stable} with respect to local perturbations. In
order to ascertain the stability of the different scaling solutions,
we should examine how localized perturbations around them evolve in
time. Since the Fokker-Planck equation is linear, the evolution of
such localized perturbations is independent of that of the scaling
solution. This simplifies the stability analysis: we need only to
solve the Fokker-Planck equation for localized initial conditions.

We now use the result (\ref{eq:ScalingFunctionBeta1}) from the exact
calculation, and find that at late times, \eqn
(\ref{eq:LocalDisturbance}) evolves to
\begin{equation}
P^*(x) t^{-\beta} f_\beta C\Bigl( \frac{x}{\sqrt{t}}\Bigr) + \delta
P(x,t) \approx  P^*(x) \Bigl[ C
t^{-\beta}f_\beta\Bigl(\frac{x}{\sqrt{t}}\Bigr) + \tilde{C}t^{-1}
f_1\Bigl(\frac{x}{\sqrt{t}}\Bigr)\Bigr].
\end{equation}
When $\beta < 1$ the second term on the rhs is negligible compared
to the first, and the scaling solution is stable to localized
perturbations. On the other hand, when $\beta > 1$ the second term
dominates the late time behavior. Such scaling solutions are
unstable and can never be observed in physical systems. We thus see
that initial conditions of the form (\ref{eq:InitialAsymptotics})
with $a < 2$, which ``excite'' scaling solutions $f_\beta$ with
$\beta < 1$, evolve to a scaling solution which depends on $a$.
``Most'' initial conditions, however, evolve towards the
\emph{marginally stable} scaling solution of $\beta = 1$. By
``most'' we mean that the basin of attraction of the  $\beta = 1$
solution in the space of all initial conditions has a higher
dimension than the basins of attraction of solutions with any $\beta
< 1$. Note that when $a>2$, the constant $\tilde{C}$ is determined
by the localized perturbation rather than the tail of the initial
condition, and therefore it is not given by \eqn
(\ref{eq:FinalConstant}).

\subsection{Scaling solution for $|x|\ll \sqrt{t}$ (``small $x$'')}\label{sec:ScalingSmallX}

In this section we show that when $|x| \ll \sqrt{t}$, the
probability $P(x,t)$ is also given in the long time limit by a
scaling form. This scaling form is different from the one discussed
above, but it, too, depends on the initial condition. Surprisingly,
this scaling form is universal: it depends on the full details of
the potential $V(x)$ only through the stationary distribution
(\ref{eq:EqDist}) which multiplies the scaling function.

Unlike the large-$x$ scaling form, the small-$x$ scaling form
depends on the boundary condition at the origin. As mentioned above,
we assume in this section that the probability current at the origin
(defined in \eqn (\ref{eq:FokkerPlanck})) vanishes at all times:
$J(0,t) = 0$. This boundary condition translates into
\begin{equation}\label{eq:SmallXBC}
\frac{\partial G}{\partial x}\biggr|_{x=0,t} = 0
\end{equation}
as long as $P^*(0) \neq 0$. The results for other boundary
conditions are discussed in \sect \ref{sec:NonConservingBC}.

From the calculation of \sect \ref{sec:ScalingLargeX} we already
know that at $x$'s which scale as $\sqrt{t}$, $G(x,t)$ is given by
(\ref{eq:ScalingAnsatz}) and (\ref{eq:FinalScalingFunction}) with
$\beta$ which depends on the initial condition. We now examine the
solution at $x$ which scale as $t^{-\gamma}$, with
$0\leq\gamma<\frac{1}{2}$. To this end, we look for scaling
solutions of the form
\begin{equation}\label{eq:SmallXScalingAnsatz}
G(x,t) = t^{-\delta_\gamma}g_\gamma\Bigl(\frac{x}{t^{\gamma}}\Bigr),
\end{equation}
which we call ``the solution at scale $t^\gamma$''.

We begin by considering the unscaled solution $G(x,t)$ itself (this
is the case $\gamma=0$). Substituting the ansatz
\begin{equation}\label{eq:SmallXScalingAnsatz0}
G(x,t) = t^{-\delta_0}g_0(x),
\end{equation}
in the Fokker-Planck equation (\ref{eq:FokkerPlanckForG}) yields
\begin{equation}
g_0''(x) - V'(x) g_0'(x) = -\delta_0 t^{-1}g_0(x).
\end{equation}
For $t\gg 1$ the term on the rhs becomes negligible.\footnote{More
precisely, we expand $G(x,t)$ in a power series in $t^{-1}$: $G(x,t)
= C_3 t^{-\delta}[a_0(x) + a_1(x)t^{-1} + a_2(x)t^{-2} +  \ldots]$,
which we substitute in (\ref{eq:FokkerPlanckForG}) and solve
separately at each order. As is shown below, at the zeroth order we
find $a_0(x) = 1$. The next order gives $a_1(x) = C' -\delta
\int_0^x dy \int_0^y dz \exp(V(y)-V(z))$, which for $x\gg 1$ is
approximately $a_1(x)\sim x^{b+1}$. This means that as long as $x\ll
t^{1/(b+1)}$ the approximation $g_0(x) = C_3$ is valid.} We thus
arrive at the simple equation $g_0''(x) - V'(x) g_0'(x) = 0$ which
can be integrated, yielding
\begin{equation}\label{eq:smallXScalingSolution0}
g_0(x) = C_3 + C_4 \int_0^x e^{V(y)}dy,
\end{equation}
where $C_3$ and $C_4$ are integration constants. The boundary
condition (\ref{eq:SmallXBC}) implies that $g_0'(0)=0$, which means
that $C_4 = 0$. Therefore, for values of $x$ which are small enough,
$G(x,t) = C_3 t^{-\delta_0}$.

We now proceed to examine the solution at scales $t^\gamma$ with
$0<\gamma<\frac{1}{2}$. At late times, $x \sim t^\gamma \gg 1$, and
we can replace $V(x) \approx b\log x$. Substituting the ansatz
(\ref{eq:SmallXScalingAnsatz}) in the equation
(\ref{eq:FokkerPlanckForGLargeX}) yields the ordinary differential
equation in the scaling variable $z = xt^{-\gamma}$
\begin{equation}\label{eq:SmallXScalingODEFull}
g_\gamma''(z) - \frac{b}{z}g_\gamma'(z) = -[\gamma z g_\gamma'(z) +
\delta_\gamma g_\gamma(z)]t^{-(1-2\gamma)}.
\end{equation}
As before, we assume that the two terms which are proportional to
$t^{-(1-2\gamma)}$ are negligible at large times and we drop them.
The validity of this assumption will be examined below. We are left
with the equation:
\begin{equation}\label{eq:SmallXScalingODE}
g_\gamma''(z) - \frac{b}{z}g_\gamma'(z) = 0,
\end{equation}
whose solution is given by
\begin{equation}\label{eq:smallXScalingSolution}
g_\gamma(z) = C_5 + C_6 z^{b+1}.
\end{equation}
The picture that emerges is that at every scale $t^\gamma$ the
solution is either a constant $C_5$ or a power law $C_6 z^{b+1}$. At
a single intermediate scale, both $C_5 \neq 0$ and $C_6 \neq 0$.
Continuity at small $x$ implies that if $C_5 \neq 0$ then $C_5 =
C_3$. With an abuse of notation, we shall from now on denote the
exponent of this special intermediate scale by $\gamma$. At this
intermediate scale, we expect the solution for large values of $z$
to coincide with the small $u$ behavior of the solution in the $x
\sim \sqrt{t}$ region (see \eqns
(\ref{eq:FinalScalingAnsatz})--(\ref{eq:FinalBeta})). This yields
the condition (\ref{eq:ODEBoundaryCondition}). Using $G(ut^{1/2},t)
\approx C t^{-\beta(a)} u^{b+1}$ (see (\ref{eq:AsymptoticF})), we
find that $C_6 = C$, and
\begin{equation}\label{eq:SmallXDelta}
\delta_\gamma(a) = \beta(a) + (b+1)({\textstyle
\frac{1}{2}}-\gamma).
\end{equation}
We note that for the scaling solution
(\ref{eq:smallXScalingSolution}), the two terms that were neglected
when passing from \eqn (\ref{eq:SmallXScalingODEFull}) to
(\ref{eq:SmallXScalingODE}) are indeed negligible as long as $z \ll
t^{\frac{1}{2}-\gamma}$, or equivalently $x \ll t^\frac{1}{2}$.

We are now left with the problem of ascertaining the values of the
two remaining undetermined constants $\gamma$ and $C_3$. These can
be found with the help of the conservation of probability (we once
again make use of the boundary condition (\ref{eq:SmallXBC})).
Choosing $t^\gamma \ll x_1(t) \ll t^{1/2}$, we can write
\begin{equation}\label{eq:SmallXProbConservation}
0 = \int_0^\infty \bigl[P(x,t)-P^*(x)\bigr]dx = \int_0^\infty
P^*(x)G(x,t) dx = I_1(t) + I_2(t),
\end{equation}
where we have defined
\begin{eqnarray}
I_1(t) &\equiv \int_0^{x_1(t)}P^*(x)G(x,t)dx \approx
t^{-\delta_\gamma}\int_0^{x_1(t)}P^*(x)\bigl[C_3 +
Cx^{b+1}t^{-\gamma(b+1)}\bigr]dx, \nonumber \\
I_2(t) &\equiv \int_{x_1(t)}^\infty P^*(x)G(x,t)dx \approx
t^{-\beta}\int_{x_1(t)}^\infty \frac{C}{Z} x^{-b}
f_{\beta}\Bigl(\frac{x}{\sqrt{t}}\Bigr)dx.
\end{eqnarray}
In these equations we have substituted the small-$x$ and large-$x$
scaling solutions for $G$. To leading order in $t$, \eqn
(\ref{eq:SmallXProbConservation}) gives $C_3 t^{-\delta_\gamma} =
-\frac{2C}{rZ} \int_0^\infty
u^{-b}f_\beta(u)du\,t^{-\beta-(b-1)/2}$, where
\begin{equation}
r=2\int_0^\infty P^*(x) dx =\left\{
\begin{array}{ll}
2 & \mbox{for diffusion on positive half-line} \\
1 & \mbox{for diffusion on entire real line}
\end{array}
\right..
\end{equation}
Using (\ref{eq:SmallXDelta}) we deduce that
\begin{equation}\label{eq:SmallXGamma}
\gamma = \frac{1}{b+1}
\end{equation}
and
\begin{equation}\label{eq:SmallXConstant1}
C_3 = -\frac{2C}{rZ}\int_0^\infty u^{-b}f_{\beta}(u)du =
-\frac{C}{rZ}\cdot\frac{4(b+1)}{2\beta(a)+b-1}.
\end{equation}

To sum up, we see that for any $x_1(t)$ in the range
(\ref{eq:FrontBoundaryRange}),
\begin{equation}\label{eq:FinalSmallXAnsatz}
G(|x|\leq x_1(t),t) \approx C t^{-\beta - \frac{b-1}{2}} g_{\beta}
\bigg(\frac{|x|}{t^{1/(b+1)}}\biggr),
\end{equation}
where
\begin{equation}\label{eq:FinalSmallXScalingFunction2}
g_\beta(z) = - {\textstyle \frac{4(b+1)}{rZ(2\beta+b-1)}} + z^{b+1}.
\end{equation}
Once again we find a scaling solution which depends on (the tails
of) the initial condition. On the other hand, just like the scaling
solution at large-$x$, this solution is essentially independent of
the full details of the potential $V(x)$, which only serves to
determine the stationary solution $P^*(x)$ and, when $a>2,$ the
constant $C$.

We emphasize that the analysis presented above holds for all
$|x|<x_1(t)$, including the region around the origin where the
potential is not logarithmic. For any fixed $x$ (which does not
scale with $t$), \eqns (\ref{eq:FinalSmallXAnsatz}) and
(\ref{eq:FinalSmallXScalingFunction2}) agree with rigorous results
obtained for discrete random walks in a logarithmic potential
\cite{MenshikovPopov1995RW,Alexander2011BesselLikeRWs}.

\subsection{Universality of late-time scaling
solutions}\label{sec:universality}

In this section, we establish the universality of the results of
\sect \ref{sec:ScalingLargeX}. That is, we show that they depend
only on the logarithmic tail of the potential, and not on the $h(x)$
correction term of \eqns (\ref{eq:ForceCorrection}) and
(\ref{eq:FokkerPlanckForG}). Moreover, our argument demonstrates
that the details of the initial condition near the origin are also
irrelevant for the large-$x$ scaling form.

To establish the required universality, we rescale $x$, $t$ and
$G(x,t)$ by defining a rescaled function
\begin{equation}\label{eq:RescaledSolutionDef}
G_\myL(x,t) \equiv \myL^{2\beta}G(\myL x,\myL^{2}t).
\end{equation}
Thus, up to a normalization factor which depends on $\beta$, the
rescaled function $G_\myL(x,t)$ is equal to $G$ at time $\myL^{2}t$
as seen at the spatial scale $\myL x$. The equation for the
evolution of $G_\myL$ may be straightforwardly obtained by
substituting the definition (\ref{eq:RescaledSolutionDef}) into \eqn
(\ref{eq:FokkerPlanckForG}), yielding
\begin{equation}\label{eq:RescaledFokkerPlanck}
\frac{\partial G_\myL(x,t)}{\partial t} =
-\frac{b}{x}\bigl(1+h_\myL(x)\bigr) \frac{\partial
G_\myL(x,t)}{\partial x} + \frac{\partial^2 G_\myL(x,t)}{\partial
x^2},
\end{equation}
where $h_\myL(x) \equiv h(\myL x)$.

The solution $G(x,t)$ at a given late time $t \gg 1$ can be obtained
in two ways: either by propagating the initial condition according
to \eqn (\ref{eq:FokkerPlanckForG}), or by rescaling the initial
condition, propagating it according to \eqn
(\ref{eq:RescaledFokkerPlanck}) to (rescaled) time 1 and rescaling
back. In the second way, the correction $h(x)$ is negligibly small.
According to this procedure, we obtain the scaling limit by
replacing $\myL$ with $\sqrt{t}$:
\begin{equation}\label{eq:ScalingLimit}
t^\beta G(u t^{1/2},t) = G_{\sqrt{t}}(u,1) \underset{t\to
\infty}{\longrightarrow} G_\infty(u,1) \equiv {f}(u).
\end{equation}
The exponent $\beta$ must be chosen appropriately so that the
$t\to\infty$ limit exists and is not zero. The limiting function
$G_\infty(x,t)$ evolves according to (\ref{eq:RescaledFokkerPlanck})
with
\begin{equation}
h_\infty(x) = \lim_{\myL \to\infty}h_\myL (x) \sim \lim_{\myL
\to\infty} (\myL x)^{-\sigma} = 0
\end{equation}
for any $x\neq 0$, see (\ref{eq:ForceCorrection}). Therefore, the
scaling limit of the original Fokker-Planck equation does not depend
on $h(x)$.

The initial condition for the rescaled problem
(\ref{eq:RescaledFokkerPlanck}) is
\begin{equation}\label{eq:RescaledInitial}
G_\infty(x,0) \equiv \lim_{\myL \to\infty}G_{\myL }(x,0)  =
\lim_{\myL \to\infty} \myL^{2\beta}G(\myL x,0).
\end{equation}
This limiting initial condition $G_\infty(x,0)$ is in many cases a
singular function, similar to the Dirac $\delta$-function but with a
different type of singularity. If the tails of $G(x,0)$ at $x \to
\pm\infty$ decay with $|x|$ faster than algebraically (e.g.,
exponentially), then $G_{\infty}(x,0)$ is zero when $x\neq 0$ and is
singular at the origin.\footnote{Consider for example a symmetric
localized initial condition $G(x,0)$ which is negative at the
origin, becomes positive at $|x| = 1$, and is exactly zero for
$|x|\geq 2$. In this case $G_{\myL}(x,0)$ is somewhat similar to
$\delta''(x)$, the second derivative of the Dirac $\delta$ function,
but might be either more or less singular than $\delta''(x)$: if
$\phi(x)$ is a smooth test function, then $\int \phi(x)
G_{\myL}(x,0) \exp[{-V_{\myL}(x)}]dx \sim
\phi''(0)\myL^{2\beta+b-3}$, which in the limit $\myL \to\infty$
might diverge or vanish, depending on the sign of $2\beta + b-3$.
Here, $V_\myL$ is the rescaled potential, defined by $V_\myL'(x) =
\frac{b}{x}(1+h_\myL(x))$.} The exact details of the initial
condition around the origin are lost in the limit which yields
$G_\infty(x,0)$. The \emph{tails} of the initial condition may,
however, affect $G_\infty$: an initial condition which decays
algebraically as in \eqn (\ref{eq:InitialAsymptotics}) is rescaled
to $G_\myL(x,0) \sim \myL^{2\beta-a}A|x|^{-a}$. If $\beta = a/2$,
the rescaled initial condition $G_\infty$ has the same algebraic
decay as $G$. We see that initial conditions may affect the scaling
solution only through their tails, and that for initial conditions
with power-law tails, a limit of (\ref{eq:RescaledInitial}) exists
only when choosing $\beta \leq a/2$ (compare with
(\ref{eq:FinalBeta})).

The rescaling argument which we have presented in this section is
inspired by the renormalization group (RG) techniques used by
Goldenfeld \textit{et al.} \cite{GoldenfeldBook} and by Bricmont and
Kupiainen \cite{BricmontKupiainen} to analyze nonlinear partial
differential equations. Here we have used their method to analyze a
linear equation which is inhomogeneous in space. From the RG
perspective, the rescaling transformation
(\ref{eq:RescaledSolutionDef}) can be viewed as an RG transformation
that has a one-parameter family of fixed points $f_\beta(u)$. The
scaling limit of the original equation is determined by the
appropriate fixed point, which is not affected by the addition of
$h(x)$. Therefore, the $h(x)$ term in the equation is irrelevant and
the scaling solution is universal in the RG sense.

\section{Comments on the scaling
solution}\label{sec:CommentsOnScaling}

In this section we comment on the scaling solution derived above and
discuss it in some broader contexts.

\subsection{The scaling solution and incomplete
self-similarity}\label{sec:IncompleteSelfSimilarity}

The dependence of the scaling exponent $\beta$ on the initial
condition signals the failure of dimensional analysis. The latter is
easily seen to predict incorrectly that $\beta = \frac{1-b}{2}$. The
reason for this failure is the following. The prediction of
dimensional analysis for the diffusion equation rests crucially on
the conservation of probability \cite{BarenblattBook}. When $b>1$,
the limiting rescaled equation (\ref{eq:FokkerPlanckLargeX}) does
not conserve probability at the origin because of the singularity of
the potential there
\cite{FirstPassageBesselProcess2011,FogedbyFiniteTimeSingularities2002}.
It is the $U(x)$ term in the potential (\ref{eq:LogPotentialFull})
which guarantees conservation of probability, and rescaling it away
(as was done in \sect \ref{sec:universality}) yields a singular
limit. In practice, this means that at any finite time $t$, no
matter how late, corrections due to the potential $U(x)$ inevitably
affect the form of the solution at small enough values of $x$. The
scaling form (\ref{eq:ScalingAnsatzForP}), in which these
corrections are not taken into account, should not be expected to
conserve probability by itself, and therefore, the scaling exponent
$\beta$ cannot be found by dimensional analysis. In the terminology
of Barenblatt, the scaling solution to our problem exhibits
self-similarity of the second kind (see \cite{BarenblattBook}).

It is interesting to note that when $b\leq 1$, the solution of the
diffusion equation (\ref{eq:LogPotential})--(\ref{eq:FokkerPlanck})
approaches a self-similar solution of the first kind, i.e., one
whose scaling exponents \emph{can} be determined by dimensional
analysis. The scaling solution in this case was found in
\cite{ProbeParticles,BarkaiKessler}. Since there is no equilibrium
distribution when $b\leq 1$, one cannot define $G(x,t)$ according to
\eqn (\ref{eq:GDefinition}). Nonetheless, one may look for scaling
solutions of the form
\begin{equation}\label{eq:ScalingAnsatzForPWrong}
P(x,t) \sim x^{-b} t^{-\beta}
\tilde{f}\Bigl(\frac{x}{\sqrt{t}}\Bigr).
\end{equation}
Dimensional analysis (or, equivalently, the conservation of
probability) dictates as before that $\beta = \frac{1-b}{2}$, which
in this case is indeed the correct value, regardless of the initial
condition. For example, when $b=0$, i.e., in the simple case of free
diffusion, one obtains the well known result $\beta = \frac{1}{2}$.

\subsection{Selection and propagating fronts}\label{sec:SelectionFronts}

A selection mechanism similar to the one described in \sect
\ref{sec:selection}, by which most initial conditions evolve into a
marginally-stable state, is well known to exist in several other
problems
\cite{VanSaarloosReview2003,MukamelSelectionLesHouches,MajumdarSelectionReview2003,Meerson1998ORipening}.
Many of these problems can be expressed as propagation of fronts
into unstable states \cite{VanSaarloosReview2003}. A well-studied
example is given by the non-linear diffusion equation which was
studied originally by Kolmogorov, Petrovsky and Piskunov
\cite{KPP1937} and by Fisher \cite{Fisher1937}:
\begin{equation}\label{eq:fkpp}
\frac{\partial \phi}{\partial t} = \frac{\partial^2 \phi}{\partial
x^2} + \phi - \phi^3.
\end{equation}
Their original works concern the spreading in space of an
advantageous mutation in a population. In this context
$0\leq\phi(x,t)\leq 1$ describes the fraction of individuals located
at point $x$ who posses an advantageous gene. This equation admits
two stationary homogeneous solutions: an unstable solution $\phi(x)
= 0$ and a stable solution $\phi(x) = 1$. Any localized initial
perturbation around the $\phi = 0$ state grows into two traveling
waves propagating outwards with an asymptotically constant velocity.
This velocity of front propagation cannot, however, be easily
determined, as \eqn (\ref{eq:fkpp}) has a traveling wave solutions
$\phi(x,t) = f_v(x-v t)$ for every possible velocity $v$.

The selection mechanism for the problem of propagating fronts has
strong similarities to our problem of diffusion in a logarithmic
potential. It is possible to show \cite{VanSaarloosReview2003} that,
similarly to our problem, the selected front solution depends on the
tails of the initial condition: if $\phi(x,0) \sim e^{-\lambda x}$,
then the asymptotic velocity is $v(\lambda) = \lambda + 1/\lambda$
for $\lambda < \lambda^* = 1$, and is $v(\lambda^*) = 2$ independent
of $\lambda$ for steep enough initial conditions, i.e., when
$\lambda > \lambda^*$ (the latter case includes localized initial
conditions, i.e., those which have a compact support). Moreover, all
traveling wave solutions with $v<v(\lambda^*)$ are unstable to
small, localized disturbances. Notice also that both in our problem
and in the problem of front propagation, stable solutions decay
monotonically at the tails, while unstable solutions decay at the
tails through oscillations (in our problem, this is a property of
the hypergeometric function (\ref{eq:FinalScalingFunction}); for
propagating fronts see, e.g., \cite{VanSaarloosReview2003}). The
marginally stable solution, into which localized initial conditions
evolve, is the solution with the steepest tail which is still
monotonic.

The similarity between our problem and the selection of propagating
fronts is furthered by noticing that, by a simple change of
variables, scaling solutions in general can be thought as traveling
waves \cite{BarenblattBook}: by defining $\xi = \log x$ and $\tau =
\log t$, any scaling solution can be expressed as\footnote{This
transformation is only valid for $x>0$. One can separately transform
the negative $x$ scaling form into a traveling wave solution by
defining $\xi' = \log(-x)$.}
\begin{equation}\label{eq:ScalingToTravelingWave}
t^{-\beta} f\Bigl(\frac{x}{t^v}\Bigr) = e^{-\beta \tau}
f\bigl(e^{\xi - v \tau}\bigr) \equiv e^{-\beta \tau} \phi(\xi - v
\tau),
\end{equation}
which is a traveling wave solution (whose overall height might
shrink or expand with time, depending on the sign of $\beta$). Note
also that a power law tail of the initial conditions
(\ref{eq:InitialAsymptotics}) implies an exponential tail in the
traveling wave variables: $A x^{-a} = A e^{-a\xi}$.

A few peculiarities of the selection problem posed by \eqn
(\ref{eq:FokkerPlanck}) should be mentioned. First, unlike in the
problem of propagating fronts, the velocity of the traveling wave
(\ref{eq:ScalingToTravelingWave}) which corresponds to the fast
front solution (\ref{eq:FinalScalingAnsatz}) and
(\ref{eq:FinalBeta}) is independent of the selected solution: it is
always $v=1/2$. Instead, it is the exponent $\beta$ which is
selected by the initial condition. In addition, as discussed above,
localized initial distributions for the diffusion equation
correspond to initial conditions (\ref{eq:InitialAsymptotics}) with
$a=0$, which, in the context of selection, are not localized (in
other words, when the distribution $P(x,0)$ is localized, then
$G(x,0)$ is not localized, and $\beta=0$ is selected rather than the
marginal value 1). This is in contrast with the many problems of
selection in which the generic initial conditions which are natural
to consider are the localized ones. Furthermore, other ``non-steep''
initial conditions are physically relevant in many situations, as
will be discussed in \sect \ref{sec:applications}. In other words,
unlike many other selection problems, \eqn (\ref{eq:FokkerPlanck})
naturally leads us to study those cases in which the solution
\emph{does} depend on the initial condition (another such exception
is found in \cite{DerridaSpohn1988}).

Another difference of the diffusion problem from most known problems
of selection lies in the fact that the scaling solution of the
diffusion problem is made up of \emph{two} scaling functions. As
mentioned above, the scaling form of the slower front is determined
by that of the faster one, and hence it is also selected by the
initial condition. In the language of traveling waves, this
corresponds to a case in which two moving fronts exist, propagating
at different velocities. While there are systems which are known to
develop two fronts selected by a marginal stability mechanism, we
are not aware of a case in which the velocities of the two fronts
are related to each other by an expression akin to \eqn
(\ref{eq:FinalDelta}).

An interesting feature of \eqn (\ref{eq:FokkerPlanck}) is that,
unlike other problems where selection takes place, this equation is
linear, yet not homogeneous in space. The linearity of \eqn
(\ref{eq:FokkerPlanck}) enables the derivation of an exact solution
(as is done in \ref{sec:calculation}), and thus assists in analyzing
the selection mechanism in detail. It should be noted that while
many problems of selection are conjectured to be governed by a
marginal stability criterion, a rigorous proof of this fact is
rarely known. The simpler linear example provided by \eqn
(\ref{eq:FokkerPlanck}) might help to shed light on the common
mathematical structure governing these similar problems.

\subsection{Relation with previous results}\label{sec:PreviousWork}
We briefly comment on the relation of our results to those of
\cite{GodrecheLuck2001ZetaUrn,ProbeParticles,BarkaiKessler}. There,
a scaling solution of the form (\ref{eq:ScalingAnsatzForPWrong})
rather than (\ref{eq:ScalingAnsatz}) was sought. For such a scaling
solution, $\beta$ must be equal to zero and
\begin{equation}\label{eq:ODEBCWrong}
\tilde{f}(u\ll 1) = 1/Z + O(u)
\end{equation}
must hold, since $P(x,t)$ should eventually converge to the steady
state distribution (\ref{eq:EqDist}). The scaling function
$\tilde{f}$ satisfies the same differential equation
(\ref{eq:ScalingODE}) as $f$, whose solution is
(\ref{eq:ODESolution}). Using the asymptotics of the hypergeometric
function (\ref{eq:1F1Asymptotic}), the boundary condition
(\ref{eq:ODEBCWrong}) together with $\tilde{f}(u\to \infty)\to 0$
determine the constants $C_1$ and $C_2$, and the scaling solution is
found to be $\tilde{f}(u) = \Gamma(\frac{b+1}{2},\frac{u^2}{4}) / Z
\Gamma(\frac{b+1}{2})$, where $\Gamma(a,x)$ is the incomplete
$\Gamma$-function.

Examining the general solution (\ref{eq:FinalScalingAnsatz}),
(\ref{eq:FinalScalingFunction}), (\ref{eq:FinalBeta}), and
(\ref{eq:FinalConstant}), and using properties of the hypergeometric
functions \cite{AbramowitzStegun}, it can be verified that when
$a=0$ and $A=-1$, the scaling exponent $\beta$ indeed equals zero
and the large-$x$ scaling solution reduces to the results of
\cite{GodrecheLuck2001ZetaUrn,ProbeParticles,BarkaiKessler}. This
case includes the large class of initial conditions $P(x,0)$ which
decay to zero faster than a power law, e.g., $P(x,0) =
\delta(x-x_0)$ (see Table \ref{tab:InitialConditions}). Other
initial conditions, however, select different values of $\beta$ and
lead to a scaling function different from the one considered
previously.

\section{Non-conserving boundary conditions}\label{sec:NonConservingBC}
So far, we have concentrated on solutions of \eqn
(\ref{eq:FokkerPlanck}) with no-flux boundary conditions at the
origin, i.e., we have assumed that the current of probability at the
origin $J(0,t)$ is zero at all times. In this section, we describe
what happens for $J(0,t) \neq 0$. Such a situation arises in two
different scenarios: (i) the distribution $P(x,t)$ is defined only
for $x\geq 0$ and the boundary condition at the origin allows
$J(0,t) \neq 0$; (ii) $x$ is unbounded, but the initial condition is
asymptotically non-symmetric, i.e., $G(x\to\pm\infty,0) \sim A_\pm
x^{-a_\pm}$ with $a_+ \neq a_-$ or $A_+ \neq A_-$. We focus here on
the first scenario, and only briefly describe what happens in the
second.

For concreteness, we discuss a specific choice of boundary condition
at the origin: an absorbing boundary condition, i.e. $P(0,t) = 0$.
This boundary condition arises naturally in many physical problems,
especially when studying first-passage properties of the dynamics
(see \sect \ref{sec:ApplicationDNA}). Other boundary conditions
(e.g., $P(0,t) = P_0$ where $P_0$ is a constant) can be treated in a
similar manner. We remark that diffusion on the half line $x\geq 0$
with an absorbing boundary at the origin is equivalent to diffusion
on the entire real line with an initial condition which is
antisymmetric. This suggests that the case of an absorbing boundary
can be treated similarly to the unbounded $x$ which we have
considered in previous sections. We do not follow this alternative
route below, as we seek a derivation which can easily be generalized
to other boundary conditions.

When probability is not conserved at the origin, the eventual steady
state which the system reaches need not be $P^*(x)$ (which by
definition (\ref{eq:EqDist}) is normalized to 1). Thus, we need to
redefine $G(x,t)$, as we expect to find a scaling form for solutions
which eventually relax to zero. We therefore define
\begin{equation}
P_\infty(x) \equiv \lim_{t\to \infty} P(x,t)
\end{equation}
to be the steady state which the system eventually reaches, and
generalize the definition of $G$ to
\begin{equation}\label{eq:GDefinitionGeneralized}
G(x,t) = \frac{P(x,t)-P_\infty(x)}{P^*(x)}
\end{equation}
(compare with (\ref{eq:GDefinition})). Note that $P_\infty(x)$
depends on the boundary condition at the origin. For instance, for a
reflecting boundary condition $P_\infty(x) = P^*(x)$, while an
absorbing boundary results in $P_\infty(x) = 0$. The definition
(\ref{eq:GDefinitionGeneralized}) allows us to consider both cases
on the same footing.

The parameter $a$ is defined by the tails of $G(x,0)$, which depends
by definition on the boundary condition (see \eqn
(\ref{eq:GDefinitionGeneralized})). Therefore, the same initial
condition $P(x,0)$ may result in two different values of $a$ when
considering two different boundary conditions (conversely, one may
say that for different boundary conditions, the same initial
condition $G(x,0)$ corresponds to different initial distributions
$P(x,0)$). A few examples of different initial conditions and the
corresponding values of $a$ and $A$ in the case of an absorbing
boundary are presented in Table
\ref{tab:InitialConditionsAbsorbing}.

\begin{table}[t!]
  \begin{center}
  \begin{tabular}{>{$}c<{\quad{}$} >{$}r<{$} @{$\;\sim\;{}$} >{$}l<{\;{}$} >{$}c<{$} >{$}c<{$}}
    \toprule
    P(x,0) & \multicolumn{2}{>{$}c<{$}}{G(x,0)} & \; a\;{} & \;A\;{}\\
    \midrule
    \delta(x-x_0) & \frac{\delta(x-x_0)}{P^*(x)} & 0 & \infty &  \\ \addlinespace
    C e^{-|x|/x_0} & \frac{C e^{-|x|/x_0}}{P^*(x)} & CZ|x|^b e^{-|x|/x_0}  & \infty &  \\ \addlinespace
    C|x|^{-(b+1)} + \ell(x) & \frac{C|x|^{-(b+1)} + \ell(x)}{P^*(x)} & C Z|x|^{-1} & 1 & C Z \\\addlinespace
    C|x|^{-(b-1)} + \ell(x) & \frac{C|x|^{-(b-1)} + \ell(x)}{P^*(x)} & C Z|x|^{+1} & -1 & C Z \\\addlinespace
    P^*(x) + C|x|^{-(b+1)} + \ell(x) & 1+\frac{C|x|^{-(b+1)} + \ell(x)}{P^*(x)} & 1 & 0 & 1 \\\addlinespace
    P^*\bigl(|x| + x_0\bigr) + \ell(x) & \frac{P^*(|x|+x_0) + \ell(x)}{P^*(x)}  &  1 & 0 & 1 \\\addlinespace
    P^*(x)\bigl[1 + e^{-|x|/x_0}\bigr] + \ell(x)& 1 + e^{-|x|/x_0} + \frac{\ell(x)}{P^*(x)} & 1 & 0 & 1 \\\addlinespace
    C P^*(x)\bigl[1 + e^{-|x|/x_0}\bigr] & C+Ce^{-|x|/x_0} & C  & 0 & C \\
    \bottomrule
    \end{tabular}
  \caption{\figtext A variety of initial conditions $P(x,0)$ and the
  corresponding values of $a$ and $A$ according to \eqn (\ref{eq:InitialAsymptotics})
  for a system with an absorbing boundary at the origin (in which case $P_\infty(x) = 0$).
  $G_0(x)$ is defined by \eqn (\ref{eq:GDefinitionGeneralized}),
  and its leading asymptotic behavior for $|x| \gg 1$ is presented. $\ell(x)$ and
  $C$ are a compactly-supported function and a constant whose values
  change from line to line. They are added to ensure the
  normalization $\int P(x,0)dx = 1$. The equilibrium distribution
  $P^*(x)$ is given in \eqn (\ref{eq:EqDist}). In some cases, the
  values of $a$ and $A$ for the same initial condition might differ
  when the boundary condition is changed (compare with Table
  \ref{tab:InitialConditions}).
  }
  \label{tab:InitialConditionsAbsorbing}
  \end{center}
\end{table}

Examining the argument of Sections \ref{sec:ScalingLargeX} and
\ref{sec:selection}, we see that the boundary condition at the
origin does not play any role in the derivation of scaling form at
large values of $x$. We can therefore conclude that the large-$x$
scaling form is independent of the boundary condition at the origin.
This conclusion is supported by the exact calculation of
\ref{sec:calculation}.

The small-$x$ scaling function, on the other hand, does depend on
the boundary condition. As in \sect \ref{sec:ScalingSmallX}, we
start by considering an ansatz (\ref{eq:SmallXScalingAnsatz0}) for
the unscaled solution $G(x,t)$, and obtain \eqn
(\ref{eq:smallXScalingSolution0}). When the origin is absorbing, the
boundary condition on $G$ is $G(x,0) = 0$, from which we deduce that
$C_3 = 0$. We therefore have
\begin{equation}\label{eq:FinalSmallXAnsatzAbsorbing}
G(x\leq x_1(t),t) \approx C t^{-\tilde{\delta}} \tilde{g}_{\beta}
(x),
\end{equation}
where
\begin{equation}\label{eq:FinalSmallXScalingFunctionAbsorbing}
\tilde{g}_\beta(x) = \tilde{C}_4 \int_0^x e^{V(y)}dy,
\end{equation}
and $x_1(t)$ is in the range $1 \ll x_1(t) \ll \sqrt{t}$ (compare
with
(\ref{eq:FinalSmallXAnsatz})--(\ref{eq:FinalSmallXScalingFunction2})).
The constant $C$ is the same as in \eqn
(\ref{eq:FinalScalingFunction}), and $\tilde{C}_4 = C_4/C$. From
(\ref{eq:FinalSmallXScalingFunctionAbsorbing}) together with the
form (\ref{eq:LogPotential}) of the potential it is seen that
$\tilde{g}_\beta(x) \sim x^{b+1}/(b+1)$ for $x\gg 1$. Matching the
large-$x$ asymptotics of (\ref{eq:FinalSmallXAnsatzAbsorbing}) with
the small-$u$ asymptotics of (\ref{eq:FinalScalingFunction}) yields
\begin{equation}\label{eq:FinalDeltaAbsorbing}
\tilde{\delta}(a) = \beta(a) + \frac{b+1}{2}
\end{equation}
and $\tilde{C}_4 = b+1$. Note that the new form of the small-$x$
scaling function (\ref{eq:FinalSmallXScalingFunctionAbsorbing}) is
no longer independent of the small-$x$ details of the potential
$V(x)$.

\begin{figure}
  \center
  \includegraphics[width = 0.6\textwidth]{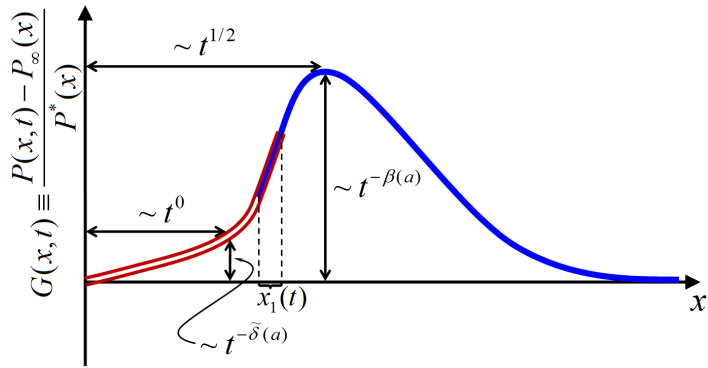}
  \caption{ \label{fig:SchematicScalingAbsorbing}\figtext{A schematic
  representation of the solution $P(x,t)$ (\eqn (\ref{eq:FinalScalingAnsatzAbsorbing}))
  at a given late time $t \gg 1$ (not drawn to scale) for a system with an absorbing boundary
  at the origin. The red hollow line represents $\tilde{g}_\beta(x)t^{-\tilde{\delta}}$,
  the solution at small values of $x$, while the blue
  solid line represents the large-$x$ scaling form
  $f_\beta(x/t^{1/2})t^{-\beta}$. The boundary condition may
  modify the value of $a$, but it does not have any further effect on the
  the large-$x$ scaling form (compare with \fig
  \ref{fig:SchematicScaling}).
  }}
\end{figure}

To sum up, changing the boundary condition at the origin affects the
scaling solution in two ways. First, it entails a change in the
definition of $G(x,0)$, which might alter the value of $a$. Second,
it modifies the small-$x$ scaling function. Importantly, the
large-$x$ scaling function and the scaling exponent $\beta$ remain
unchanged. For the case of an absorbing boundary, these changes are
summed up in the final scaling form of the solution
\begin{equation}\label{eq:FinalScalingAnsatzAbsorbing}
P(x,t) \approx P_\infty(x) + C P^*(x)\cdot \left\{
\begin{array}{ll} \tilde{g}_{\beta(a)}(x)
t^{-\tilde{\delta}(a)}  & \mbox{ for } x\leq x_1(t) \\ \\
f_{\beta(a)} \bigl(\frac{x}{t^{1/2}}\bigr)t^{-\beta(a)} & \mbox{ for
} x\geq x_1(t)
\end{array} \right.,
\end{equation}
where, $\tilde{g}$ is given in \eqn
(\ref{eq:FinalSmallXScalingFunctionAbsorbing}) and
$\tilde{\delta}(a)$ by (\ref{eq:FinalDeltaAbsorbing}). This solution
is depicted schematically in \fig
\ref{fig:SchematicScalingAbsorbing}.

Finally, we briefly comment on diffusion on the entire real axis
with non-symmetric initial conditions $G(x\to\pm\infty,0) \sim A_\pm
x^{-a_\pm}$ (scenario (ii) above). In this case, there are two
different ``fast fronts'' which propagate from the origin to
$\pm\infty$: a scaling function for $x \sim \sqrt{t}$ and another
for $x \sim -\sqrt{t}$. Each of these is selected by the
corresponding tail of the initial condition. Similarly, there are
two ``slow fronts'' (i.e., small-$x$ scaling functions), each one
overlapping with the corresponding large-$x$ scaling function. A
calculation similar to that of \sect \ref{sec:ScalingSmallX} can be
repeated, leading to solutions of the form
(\ref{eq:SmallXScalingAnsatz}), (\ref{eq:smallXScalingSolution}) and
(\ref{eq:SmallXDelta}). Four unknown variables remain: $\gamma_\pm$
and $C_{5,\pm}$ (the values of $\gamma$ and $C_5$ for the
positive-$x$ and negative-$x$ scaling functions). These can in
principle be determined from two equations: the conservation of
total probability, and continuity of the probability current at
$x=0$.

\section{Applications\label{sec:applications}}
In this section, we present applications of the new theoretical
results which have been derived above. In particular, we give
examples of several problems in which the dependence of the scaling
form on the initial conditions plays an important role.

As discussed above, when probability is conserved, a large class of
initial distributions (including localized ones) correspond to a
value of $a=0$ (see \eqns (\ref{eq:GDefinition}) and
(\ref{eq:InitialAsymptotics}) and Table
\ref{tab:InitialConditions}). For these initial conditions the
distribution evolves to the $\beta = 0$ scaling form, which is the
one previously obtained in
\cite{GodrecheLuck2001ZetaUrn,ProbeParticles,BarkaiKessler}. An
inspection of \eqns (\ref{eq:GDefinition}) and
(\ref{eq:InitialAsymptotics}) and of Table
\ref{tab:InitialConditions} reveals that initial conditions with
$a\neq 0$ can be divided into two broad classes: for $a\leq 0$, the
exponent $a$ yields the leading decay of the tail of the initial
distribution. On the other hand, when $a$ is positive, the tail of
the initial distribution approaches the equilibrium distribution
$P^*(x)$; in this case, the leading decay of the initial
distribution as $x\to\pm\infty$ is that of $P^*$, and $a$ determines
the sub-leading correction to $P^*$. Below we consider examples of
both classes. In \sect \ref{sec:ApplicationNegativeA}, we describe
an experimental protocol by which initial conditions with negative
values of $a$ can be obtained, and we propose a cold-atoms
experiment which, using this protocol, could measure the predicted
dependence of the relaxation on the initial condition. Initial
conditions belonging to the second class may at first sight seem
unnatural in physical circumstances, as they require fine-tuning the
initial distribution. In \sect \ref{sec:ZRP} we show that this is
not necessarily the case, and explain how initial conditions with
$a=1$ arise naturally in the calculation of current correlations in
the zero-range process, a stochastic model of particle transport.

When the boundary condition at the origin is absorbing, on the other
hand, the value of $a$ is always determined by the leading decay of
the tail, no matter what the initial distribution is (see \eqn
(\ref{eq:GDefinitionGeneralized}) and Table
\ref{tab:InitialConditionsAbsorbing}). Therefore, no value of $a$
requires fine tuning of the initial distribution. In \sect
\ref{sec:ApplicationDNA}, we provide one example of such a system:
we explain why the dynamics of loops in a denaturating DNA molecule
is described by \eqn (\ref{eq:FokkerPlanckLargeX}) with an absorbing
boundary, and show the implications of the dependence on initial
conditions to the analysis of results of single-molecule
experiments.

For the sake of completeness, we provide in \sect \ref{sec:Review} a
review of some other systems which are described by \eqn
(\ref{eq:FokkerPlanckLargeX}), to which our results may be relevant.

\subsection{Initial conditions with $a<0$ and
atoms in optical lattices\label{sec:ApplicationNegativeA}}

Equations (\ref{eq:GDefinition}) and (\ref{eq:InitialAsymptotics})
indicate that the tail of the initial distribution for $a<0$ is of
the form
\begin{equation}\label{eq:OpticalInitial}
P(x,0) \sim A x^{-\mu}\quad \mbox{with} \quad \mu <b.
\end{equation}
Here, $a = \mu - b$. Thus, such initial distributions can be
relatively easily generated in physical situations. This observation
straightforwardly suggests a protocol by which one can observe the
dependence of the relaxation dynamics on initial conditions. For the
sake of concreteness, we present this protocol in the context of
cold atoms trapped in optical lattices, where the dependence on
initial conditions can be tested experimentally.

When cold atoms are placed in optical lattices, their momentum
performs a diffusion which, in the semi-classical regime, is of the
form (\ref{eq:FokkerPlanckLargeX}) where $x$ represents the momentum
\cite{CohenTannoudji}. In recent years, this momentum diffusion has
received both theoretical and experimental attention due to the
power-law distribution and ``anomalous'' dynamics to which it gives
rise
\cite{Zoller1996,LutzErgodicity2003,LutzTsallis2004,
OpticalTsallisExperiment2006,SagiEtal2011OpticalXDiffusionExp}.
The Fokker-Planck equation for the semi-classical probability
distribution $W(p,t)$ of an atom with momentum $p$ at time $t$ is
\begin{equation}\label{eq:OpticalMomentumFokkerPlanck}
\frac{\partial W(p,t)}{\partial t} = \frac{\partial}{\partial
p}\biggl[-F(p)W(p,t) + D(p)\frac{\partial W(p,t)}{\partial
p}\biggr],
\end{equation}
where, in appropriate units,  $F(p) = -\frac{b p}{1+ p^2} =
-\frac{b}{p} + O(p^{-3})$ is the cooling ``friction'' force, and
$D(p) = 1 + \frac{D}{1+p^2}$ is a momentum-dependent diffusion
coefficient \cite{CohenTannoudji}. The parameters $b$ and $D$ are
determined by the depth of the optical lattice, which may be
controlled in an experiment by the intensity and detuning of the
optical lattice. When $D \ll 1$ \eqn
(\ref{eq:OpticalMomentumFokkerPlanck}) is of the form
(\ref{eq:LogPotential})--(\ref{eq:FokkerPlanck}). The equation can
be brought to this form even when $D$ is not negligible, by the
standard transformation $q(p) = \int^p \sqrt{D/D(p')}dp' = p + D
\arctan(p)$ \cite{Risken}. The transformed equation reads
\begin{equation}
\frac{\partial W(q,t)}{\partial t} = \frac{\partial}{\partial
q}\biggl[-\tilde{F}(q)W(q,t) + \frac{\partial W(q,t)}{\partial
q}\biggr],
\end{equation}
where once again $\tilde{F}(q) = -\frac{b}{q} + O(q^{-3})$. For
convenience of notation, we will assume below that $D \ll 1$ and
study \eqn (\ref{eq:OpticalMomentumFokkerPlanck}).

The experimental protocol to observe the ``anomalous'' scaling
suggested by equations
(\ref{eq:FinalScalingAnsatz})--(\ref{eq:FinalBeta}) is rather
straightforward. For any given value of the parameter $b$, the
stationary distribution of momentum is given by $W^*_b(p) = Z_b^{-1}
e^{-V_b(p)} = Z_b^{-1}(1+p^2)^{-b/2}$ (where we have made the
dependence on the parameter $b$ explicit in our notation). In an
experiment, the parameter $b$ can be controlled by changing the
depth of the optical potential. The following two-step procedure
would generate an appropriate initial condition with negative $a$:
(1) a state with momentum distribution $W(p,0) =
Z_{b+a}^{-1}(1+p^2)^{-(b+a)/2}$ with some $b>1$ and $1-b<a<0$ is
prepared by setting the parameters of the experiment to a value
which corresponds to $b+a$, and allowing the system to equilibrate;
then (2) at time $t=0$ the parameters are rapidly changed from $b+a$
to $b$. Following this ``quench'', the distribution $W(p)$ or one of
its moments is measured as a function of time. For instance, if
$b+a>3$, one may measure the variance of the momentum $\langle p^2
\rangle$ (which is proportional to the mean kinetic energy of the
atom), which is predicted to decay as
\begin{eqnarray}
\langle p^2(t)\rangle - \langle p^2(\infty)\rangle = \int dp\, p^2
[P(p,t) - P^*_b(p)] \sim t^{-\frac{b+a-2}{2}} \int
u^{2-b}f_{a/2}(u)du. \nonumber \\
\end{eqnarray}
Although we have presented this experimental protocol in the context
of cold atom experiments, it could be used in many other physical
contexts as well.

We remark that unlike many cases in which fat-tail distributions
lead to an anomalous time-evolution, in the case which we discuss
here there is no requirement that any particular moment of the
initial distribution diverges. In fact, any particular moment of the
initial (or final) distribution can be guaranteed to be finite by
selecting $b$ large enough with a fixed value of $a$.

\subsection{Initial conditions with $a=1$ and
current correlations in a critical zero-range
process}\label{sec:ZRP}

Another way to generate initial conditions with $a\neq 0$ without
fine-tuning the parameters of the initial state is to prepare the
system initially in a translate of the equilibrium distribution,
i.e., $P(x,0) = P^*(x+\Delta x)$ for some $\Delta x$. In this case,
the initial condition corresponds to $a=1$ (see Table
\ref{tab:InitialConditions}). Such a situation may be realized
experimentally if it is possible to displace the confining
logarithmic potential.

In this section we present a different case in which such an initial
condition arises. The problem we shall address here is the
calculation of stationary two-time correlations of particle currents
in a zero-range process (ZRP), a stochastic model of particle
transport exhibiting real-space condensation.

In the ZRP which we consider, $N=\rho L$ particles hop on a
one-dimensional lattice of $L$ sites with periodic boundary
conditions ($\rho$ is the density of particles). The particles can
only move in one direction. The defining property of the model is
that the rate of a jump from site $i$ to $i+1$ is a function only of
the number of particles $n_i$ in the departure site. We denote this
rate by $w(n_i)$. This non-equilibrium model of interacting
particles has been studied extensively in recent years. For certain
choices of the hopping rates, the model exhibits a condensation
transition whereby, when the density is increased above a critical
density $\rho_c$, a finite fraction of all particles resides in a
single site (selected at random). For reviews of this condensation
transition and other applications of the model see
\cite{evanszrpreview,MajumdarCondensationLesHouches,Schadschneider2010Book}.

We consider a ZRP at the critical density, and examine correlations
of the current flowing across a single site. We concentrate on
hopping rates which for large $n$ have the form
\begin{equation}\label{eq:ZRPrates}
w(n) = 1 + \frac{b}{n} + O(n^{-2}).
\end{equation}
These commonly studied rates give rise to condensation when $b>2$
\cite{EvansZRPcondensation}. In the thermodynamic limit (when $L\to
\infty$), the arrival of particles into any site is a Poisson
process with rate 1 which is independent of the the process of
particles departing from the site\footnote{In a system of finite
size $L\gg 1$, the arrival process is approximately Poisson on time
scales $t \ll L$ \cite{GuptaFiniteZRP}. Therefore, the result
obtained below (\eqn \ref{eq:ZRPCurrentCorrelatorFinal}) is correct
for finite systems in the intermediate asymptotics regime of $1 \ll
t \ll L$. } \cite{GuptaFiniteZRP}. The occupation probability of the
site $P(n)$ evolves according to the master equation
\cite{ProbeParticles}
\begin{eqnarray}\label{eq:ZRPMasterEq}
\frac{\partial}{\partial t}P(n) &= P(n\!-\!1)  +
w(n\!+\!1)P(n\!+\!1) - [1+w(n)]P(n)\approx \nonumber \\
&\approx \frac{\partial}{\partial
n}\Bigl[\frac{b}{n}\bigl(1+O(n^{-1})\bigr)P(n)\Bigr] +
\frac{\partial^2 P(n)}{\partial n^2},
\end{eqnarray}
which is of the form of \eqn
(\ref{eq:FokkerPlanckDetailed})--(\ref{eq:ForceCorrection}). It is
straightforward to verify that the steady-state distribution is
\begin{equation}\label{eq:ZRPStationaryDist}
P^*(n) = \frac{1}{Z}\prod_{k=1}^n\frac{1}{w(k)} =
\frac{1}{\mathcal{Z}}\,n^{-b} \Bigl(1+O(n^{-1})\Bigr)
\end{equation}
where $Z$ is a normalization constant and $\mathcal{Z}$ is
non-universal and depends on the full form of the rates $w(n)$. For
$w(n) = 1 + b/n$, for example, it can be shown that $Z = b/(b-1)$
and $\mathcal{Z} = [(b-1)\Gamma(b)]^{-1}$.

Having presented the model, we now present the specific problem
which we wish to study, and show how the results of previous
sections can be used to solve it. Our task is to calculate the
correlation function
\begin{equation}
C(t) \equiv C_{\mathrm{in},\mathrm{out}}(t) \equiv \langle
j_{\mathrm{in}}(0) j_{\mathrm{out}}(t) \rangle - j^2, \qquad t \geq
0,
\end{equation}
where $j_{\mathrm{in}}(t)dt$ is the number of particles arriving at
the site between time $t$ and $t+dt$, $j_{\mathrm{out}}(t)dt$ is the
number of particles departing from the site during this time period,
and $j = \langle j_{\mathrm{in}}(t) \rangle  = \langle
j_{\mathrm{out}}(t) \rangle= \sum P^*(n)w(n) = 1$ is the mean
current in the steady state. Angular brackets denote an average in
the steady state. We may similarly define the correlation functions
$C_{\mathrm{in},\mathrm{in}}(t)$, $C_{\mathrm{out},\mathrm{out}}(t)$
and $C_{\mathrm{out},\mathrm{in}}(t)$, but these are all equal to
zero: both the arrival process of particles entering the site and
the departure process of particles leaving it are Poisson
processes,\footnote{It is not a trivial statement that the departure
process is a Poisson process. In the field of queueing theory, this
statement is known as Burke's theorem, see \cite{kellybook}.} and
the arrival process is independent of the departure process. To
simplify notation we shall from now on drop the subscripts and
denote $C(t) \equiv C_{\mathrm{in},\mathrm{out}}(t)$.

Although the exact steady-state distribution of the model can be
calculated for any jump rates, little is known about two-time
correlation functions such as $C(t)$, even in the steady-state. We
now show that the long time asymptotics of this correlation function
can be found using the scaling solution
(\ref{eq:FinalScalingAnsatz}) and (\ref{eq:FinalBeta}) of \eqn
(\ref{eq:ZRPMasterEq}) with $a = 1$. To this end, we note that
$\langle j_{\mathrm{in}}(0) j_{\mathrm{out}}(t) \rangle$ is given by
a product of the rate with which a particle enters the site at time
0, (which is 1) and the conditional rate with which a particle
leaves the site at time $t$ given that a particle has entered at
time zero. The latter rate depends on the (conditional) occupation
of the site at time $t$, and therefore the correlation function is
\begin{eqnarray}\label{eq:ZRPCurrentCorrelator}
C(t) &= \sum_{n,m} P^*(n) \cdot 1 \cdot \Bigl[P(m,t|n+1,0)-P^*(m)\Bigr] w(m) = \nonumber \\
&= \sum_{m=1}^\infty  \Bigl[P(m,t|P_0)-P^*(m)\Bigr] w(m),
\end{eqnarray}
where $P(m,t|n,0)$ is the conditional probability to have $m$
particles in the site at time $t$ given that there were $n$ at time
$0$, and in the last equality we have introduced the notation
\begin{equation}\label{eq:ZRPConditionalDist}
P(m,t|P_0) \equiv \sum_{n=0}^\infty P_0(n) P(m,t|n,0) \quad
\mbox{with}\quad P_0(n) \equiv P^*(n-1).
\end{equation}
For large $n$, the initial condition $P_0(n)$ satisfies
\begin{equation}
P_0(n) = P^*(n)\Bigl[1 + b n^{-1} + O(n^{-2}) \Bigr],
\end{equation}
and therefore it is of the form (\ref{eq:InitialAsymptotics}) with
$a=1$ and $A = b$.

The calculation of $C(t)$ now proceeds by substituting the
appropriate solution (\ref{eq:FinalScalingAnsatz}) in \eqn
(\ref{eq:ZRPCurrentCorrelator}) and evaluating the sum. We carry out
this calculation in \ref{sec:ZRPappendix}. This calculation turns
out to be somewhat subtle, as the leading terms in $t$ exactly
cancel out, and the decay of correlations is determined by the
next-to-leading term. We note here that the cancelation of the
leading-order terms can only be established using both the small-$x$
and large-$x$ asymptotic regimes. The result of the calculation is
\begin{equation}\label{eq:ZRPCurrentCorrelatorFinal}
C(t) \sim  \frac{\pi \Gamma(\frac{1+b}{2})} {\mathcal{Z}
2^{b}\Gamma^2\bigl(\frac{b}{2}\bigr)}\, t^{-\frac{b+a}{2}} =
\frac{\pi \Gamma(\frac{1+b}{2})} {\mathcal{Z}
2^{b}\Gamma^2\bigl(\frac{b}{2}\bigr)}\, t^{-\frac{b+1}{2}},
\end{equation}
where $\mathcal{Z}$ is defined in \eqn (\ref{eq:ZRPStationaryDist}).

\subsection{Absorbing boundary conditions and dynamics of denatured DNA loops at criticality}\label{sec:ApplicationDNA}

The analysis of \sect \ref{sec:NonConservingBC} has revealed that
initial conditions with any value of $a$ can be achieved without
fine-tuning when the boundary at the origin is absorbing (see Table
\ref{tab:InitialConditionsAbsorbing}). Absorbing boundary conditions
arise naturally when studying first-passage problems such as the
mean time it takes a diffusing particle to reach the origin from a
given initial condition (see for example \cite{RednerBook}). In this
section, we discuss one such example in the experimental context of
the dynamics of denaturing DNA molecules.

It is well known that when the double stranded DNA molecule is
heated, it undergoes a denaturation phase transition in which it
separates into two single strands. The nature of this phase
transition has been debated over the years. Many of the theoretical
studies of this transition are based on the model of Poland and
Scheraga
\cite{PolandScheraga1966a,*PolandScheraga1966b,DnaReview1985WartelBenight,DNAGotoh1983}
(for recent reviews see \cite{DNAmelting2009SpecialIssue}). These
studies model the DNA molecule as an alternating sequence of bound
segments and denatured loops, or bubbles. The bound segments are
considered rigid, with each bound pair contributing a negative
energy of $-\epsilon$ in the case of homopolymers, while the shape
of open loops may fluctuate and thus contribute to the entropy of
the molecule. The energetic cost of initializing a loop is
$\epsilon_0>0$, and the configuration of an open loop of size $\ell$
does not further affect its energy. The number of states of a long
loop of length $\ell \gg 1$ is given by the number of random walks
of $2\ell$ steps which return to their starting point:
\begin{equation}
\Omega(\ell) \sim \frac{s^\ell}{\ell^b}.
\end{equation}
Here $s$ is a geometrical constant which depends on the microscopic
details of the molecule, while the universal exponent $b$ depends
only on space dimension and on the existence of long range
interactions in the molecule such as self-avoiding interactions: for
a non-self-avoiding loop in $d$ dimensions $b = d/2$, while
self-avoiding interactions, both within the loop and between the
loop and the rest of the molecule, were shown to increase the value
of $b$ to approximately 2.11 in $d=3$ dimensions
\cite{KafriMukamelPelitiPRL,DnaCarlonEtalNumericalC2002}. The value
of this exponent has received much attention, since it determines
the order of the transition: for $1<b<2$ the transition is second
order, while $b>2$ leads to a first order transition.

In recent years, with the advent of single molecule experiments,
direct measurements of the dynamics of denatured segments became
possible
\cite{DNAKrichevsky2003BubbleDynamics,DnaAmbjornssonEtal2006Breathing}.
In particular, the state of a single tagged base pair can be
followed using fluorescence correlation spectroscopy, whereby
fluorescence occurs as long as the base pair is open and is quenched
when it is closed. Such experimental developments have lead to a
theoretical effort to study the dynamics of denaturation using the
Poland-Scheraga model
\cite{DnaHankeMetzler2003,BarKafriMukamel2007DNA,FogedbyMetzler2007DNA,
bar2009DNAMelting,DnaKunzLivi2007,SchutzMurthy2011DNA}. These
studies consider dynamics which obey detailed balance with respect
to the Poland-Scheraga free energy: if $w_\pm(\ell)$ are the rates
with which a loop of length $\ell$ changes its length by $\pm 1$,
then $w_+(\ell)/w_-(\ell+1) = s
e^{-\beta\epsilon}(\frac{\ell}{\ell+1})^b$. At the transition
temperature $T_m$, open loops are sparse. They rarely coalesce or
split up since $\epsilon_0 \simeq 10 k_B T_m$
\cite{BlosseyCarloDNACooperativity}. Therefore, to a good
approximation, the dynamics of a single loop may be considered
independently of that of other loops. From these considerations one
may conclude that at the melting temperature, the loop-length
probability distribution evolves according to a master equation
which, when the loop size is large, approaches the Fokker-Planck
equation (\ref{eq:FokkerPlanckLargeX}) (where $x$ is the loop size
$\ell$). Note that the large value of $\epsilon_0$ implies that once
the length of a loop shrinks to zero it does not reappear in the
same position for a long period of time. Therefore, an absorbing
boundary condition at $\ell=0$ is appropriate for the study of the
dynamics of denatured loops.

The fluorescence correlation function which can be measured in
experiments is related to the probability that an unbound loop
remains open after time $t$
\cite{DNAKrichevsky2003BubbleDynamics,BarKafriMukamel2007DNA,bar2009DNAMelting,FogedbyMetzler2007DNA}.
At late times, this survival probability is given by
\begin{equation}
S(t) = \sum_{\ell=1}^{\infty} P(\ell,t) = \sum_{\ell=1}^{L}
P(\ell,t) + \sum_{\ell=L+1}^{\infty} P(\ell,t) \equiv S_1(t) +
S_2(t),
\end{equation}
where $S_1(t)$ and $S_2(t)$ correspond to the contributions to the
sum from small and large loops, respectively. Here, $L\gg 1$ is a
constant. Using the scaling form
(\ref{eq:FinalScalingAnsatzAbsorbing}) for the probability
distribution, it is easy to evaluate the two sums and find that
$S_1(t) \sim t^{-\beta - (b+1)/2}$, while $S_2(t) \sim t^{-\beta -
(b-1)/2}$, where $\beta$ depends on the initial condition, as
discussed below. Therefore $S_2(t)$ dominates the sum and
\begin{equation}
S(t) \sim t^{-\beta - (b-1)/2}.
\end{equation}

When the loop is allowed to fluctuate freely, the probability of
selecting an initial loop of length $\ell_0$ in the steady state is
\begin{equation}\label{eq:DNAInitial1}
P_0(\ell_0) \sim \ell_0 P^*(\ell_0) \sim \ell_0^{-(b-1)},
\end{equation}
which is normalizable in the case of DNA where $b>2$
\cite{BarKafriMukamel2007DNA}. This is the natural
experimentally-relevant initial condition when one probes the state
of a base pair (whether it is bound or not) at random. According to
the definition of the parameter $a$ for the case of absorbing
boundary conditions, it corresponds to $a=-1$ (see \eqns
(\ref{eq:InitialAsymptotics}) and
(\ref{eq:GDefinitionGeneralized})). Thus, for the relevant initial
condition one has from \eqn (\ref{eq:FinalBeta}) $\beta = -1/2$,
yielding $S(t) \sim t^{-(b-2)/2}$, as was previously obtained in
\cite{BarKafriMukamel2007DNA} using a different method.

A different possible experimental protocol is obtained when one
forces one end of the loop to be on a particular site. In this case,
there is no need for the factor of $\ell_0$ in \eqn
(\ref{eq:DNAInitial1}) \cite{BarKafriMukamel2007DNA}. The initial
condition is then $P_0(\ell_0) = P^*(\ell_0)$, yielding $a=0$ and
$\beta = 0$. Therefore, the survival probability decays as $S(t)
\sim t^{-(b-1)/2}$, once again in accordance with
\cite{BarKafriMukamel2007DNA}.

Finally, the case of a localized initial condition, namely starting
from a loop of a given length, has been considered by
\cite{FogedbyMetzler2007DNA}. This case is far harder to realize
experimentally. In our approach, this initial condition corresponds
to $a=\infty$, which leads to $\beta =  1$ and a different behavior
of the survival probability, $S(t) \sim t^{-(b+1)/2}$.

The conclusion from this discussion is that since the initial
condition selects the value of the scaling exponent $\beta$, it may
affect all correlation functions which can be measured
experimentally. Therefore, when analyzing experiments which measure
the dynamics of denaturing DNA loops, one must carefully take into
account the appropriate initial condition which is relevant to the
experiment.

\subsection{Other systems described by \eqn (\ref{eq:FokkerPlanckLargeX})}\label{sec:Review}

In light of its simplicity, it is not surprising that \eqn
(\ref{eq:FokkerPlanckLargeX}) arises in many different contexts. We
now briefly review some of the problems described by this equation.
This review, which is far from being exhaustive, is included to
indicate the variety of problems to which the results obtained in
this paper may apply. The physical implications of our results to
these systems have so far not been worked out.

\begin{enumerate}
  \item We have considered so far only one-dimensional problems of
  diffusion in a logarithmic potential. In fact, as long as the
  problem is spherically symmetric, diffusion in a
  logarithmic potential in \emph{any} dimension leads to an equation
  of the form (\ref{eq:LogPotential})--(\ref{eq:FokkerPlanck}) for the
  diffusion in the radial direction \cite{BrayPersistence2000}. In this case,
  the parameter $b$ depends on the spatial dimension. A
  similar equation results when considering a spherically symmetric
  convection-diffusion equation in two dimensions with a sink or
  source at the origin
  \cite{KrapivskyRednerDiffusionConvection2007}.

  \item The one-dimensional diffusion equation in an attractive logarithmic
  potential can be mapped, as we show in \ref{sec:calculation},
  to the (imaginary time) Schr\"odinger equation which describes a quantum mechanical
  particle in a repulsive inverse square potential $V_s(|x|\gg 1) \sim \gamma/x^2$
  (where the coupling constant $\gamma$ is related to $b$,
  see \eqns
  (\ref{eq:ExactWavefunctionDef})--(\ref{eq:ExactSchrodingerPotential})).
  The quantum inverse square potential has drawn much attention over
  the years (for references, see for example
  \cite{QMInverseSquareAmJPhys2006,QMInverseSquareMoroz2010}).
  Although the questions we address in the present study are motivated
  by problems of diffusion, the scaling solution we have found above
  is valid also in the corresponding quantum system. It would be
  interesting to understand its implications in the context of quantum
  mechanics.

  \item Models of gases with long-range interactions exhibit slow relaxations
  towards equilibrium. One approach to study these slow relaxations is
  to examine the evolution towards equilibrium of a single tagged particle
  inside an equilibrated gas of this type. In several models it has
  been established that the kinetic equation which describes the
  relaxation of the tagged-particle momentum distribution can be
  transformed to a Fokker-Planck equation with the asymptotic form
  (\ref{eq:FokkerPlanckLargeX}), from which the time dependence of
  different correlation functions can be calculated
  \cite{BouchetDauxoisRapid2005,BouchetDauxoisLong2005} (for a review see
  \cite[\sect 5.2.3]{CampaEtalLongRangeReview2009} and references
  therein).

  \item An equation of type (\ref{eq:FokkerPlanckLargeX}) was
  encountered in the dynamics of a two dimensional $XY$ model below
  the Kosterlitz-Thouless transition. In \cite{BrayPersistence2000},
  it has been shown that this equation can describe the annihilation
  of vortex-antivortex pairs during the relaxation to equilibrium
  after a quench from high temperatures.

  \item In the study of Barkhausen noise, this equation is
  used to derive the distribution of magnetization jumps within
  the mean-field ABBM model \cite[Sect. IV, B]{BarkhauseReview2006}.

  \item In a biological context, a discrete-time version of
  \eqn (\ref{eq:FokkerPlanckLargeX}) was suggested as a model for
  the dynamics of sleep-wake transitions during a night's sleep
  \cite{LoEtAlSleepWake2002}.

  \item Many studies of \eqn (\ref{eq:FokkerPlanckLargeX}) were
  motivated not by specific physical phenomena, but by interesting
  mathematical features of the equation. These include
  studies of the persistence exponents for a diffusion described
  by \eqn (\ref{eq:FokkerPlanckLargeX}), which are
  found to depend on the dimensionless coupling constant $b$
  \cite{FirstPassageBesselProcess2011,BrayPersistence2000,FaragoPersistence2000}; an
  examination of the effect of noise on evolution equations such as
  (\ref{eq:FokkerPlanckLargeX}) which give
  rise to finite time singularities \cite{FogedbyFiniteTimeSingularities2002};
  and an examination of the relation between the tails of
  stationary distributions of Markov processes and power-law
  decay of correlations in the dynamics \cite{Miccicche2009}.
\end{enumerate}

\section{Conclusion}\label{sec:conclusion}
In this paper we considered the late-time scaling behavior of a
particle diffusing in a potential with logarithmic tails, focusing
solely on the trapping case in which the probability distribution
relaxes to a normalizable steady state. By concentrating on the
deviation from equilibrium (i.e., the difference between the
solution and the steady state), we have generalized the scaling
solution which in
\cite{GodrecheLuck2001ZetaUrn,ProbeParticles,BarkaiKessler} was
obtained for localized initial conditions to any initial condition.

The scaling solution to this, rather simple, linear diffusion
problem contains several surprises. The first is that at small
values of $|x|$, where the diffusive ($x \sim \sqrt{t}$) scaling
regime is invalid, the solution is given by a different scaling
function. Thus, to leading order in $t$, the full solution on the
entire real axis is given by the simple scaling form
(\ref{eq:FinalScalingAnsatz}). With this new result it is easy to
compute the time dependence of many correlators, even of functions
which are concentrated around the origin (e.g., $\langle 1/x(t)
\rangle$). The utility of this scaling form was demonstrated in the
calculation of current correlations in the zero-range process
presented in \sect \ref{sec:ZRP}.

Another surprising aspect of the solution is that the Fokker-Planck
equation (\ref{eq:LogPotential})--(\ref{eq:FokkerPlanck}) has
incomplete scaling solutions, i.e., solutions in which the scaling
exponents cannot be determined from dimensional analysis. Moreover,
these scaling exponents depend on the initial condition via a
selection mechanism which is similar in many of its details to the
marginal stability mechanism which governs selection in problems of
fronts propagating into an unstable state. Since our system is not
spatially homogeneous, the standard techniques which are employed in
the study of the selection of propagating fronts (most notably
Fourier analysis) are inapplicable. However, as the diffusion
equation is linear, it can be solved exactly and the selection
mechanism can be proven rigorously. We hope that the similarities
and differences between the problem we have studied here and the
selection in propagating fronts might shed light on the mathematical
structure which underlies the selection mechanism.

Beyond their intriguing and surprising mathematical properties,
these scaling solutions have considerable utility for a large
variety of physical problems which are mathematically equivalent to
the diffusion equation in a logarithmic potential. We demonstrated
the applicability of our results to three examples: an experimental
protocol was suggested, in which cold atoms in an optical lattice
are ``quenched'' from one value of the diffusion constant to
another, which should exhibit a relaxation that depends on its
initial steady-state; two-time current correlations in the
steady-state of a system undergoing a non-equilibrium real-space
condensation transition were calculated; and it was demonstrated
that initial conditions are important when analyzing experimental
data of the dynamics of denaturing DNA loops. It would be
interesting to examine how the dependence of the scaling form on the
initial condition might be manifested in other systems governed by
the diffusion equation (\ref{eq:FokkerPlanckLargeX}). A particularly
intriguing question is the significance of such scaling solutions in
the problem of a quantum mechanical particle in an inverse square
potential.

\ack  We thank A.\ Amir, A.\ Bar, O.\ Cohen, N.\ Davidson, J.-P.\
Eckmann, M.\ R.\ Evans, and T.\ Sadhu for useful discussions and
comments on the manuscript. This work was supported by the Israel
Science Foundation (ISF).

\appendix

\section{Derivation of the scaling solution}\label{sec:calculation}
In this appendix we solve exactly the Fokker-Planck equation
(\ref{eq:LogPotential})--(\ref{eq:FokkerPlanck}) and calculate its
long-time asymptotic form. We proceed by performing the calculation
only for symmetric potentials which are exactly equal to a logarithm
for large enough $x$. The scaling argument of \sect
\ref{sec:universality} implies that the long-time asymptotics we
thus obtain hold for any potential with the asymptotic form
(\ref{eq:LogPotential}). The calculation presented below is based on
the methods of \cite{Zoller1996} and \cite{FaragoPersistence2000}.
The case of $a=0$ has recently been analyzed in a similar fashion in
\cite{BarkaiKessler2011Long}.

\subsection{Mapping to a Schr\"odinger equation}

Consider a particle diffusing under the influence of a symmetric
potential
\begin{equation}\label{eq:ExactLogPotential}
V(x) = \left\{
\begin{array}{ll} \tilde{V}(x) & \mbox{for } |x|<x_0 \\
b \log(|x|) & \mbox{for } |x|>x_0.
\end{array} \right.
\end{equation}
for some $x_0>0$, where $\tilde{V}(x)$ is some symmetric potential.
By a proper rescaling of $x$ and $t$, it is always possible to set
the the threshold $x_0 = 1$. We further assume that the potential is
measured in units of temperature (i.e., $k_B T = 1$). From now on we
denote by a tilde any quantity in the region $|x| <x_0=1$.

The corresponding Fokker-Planck equation reads
\begin{equation}\label{eq:ExactFokkerPlanck}
\frac{\partial P(x,t)}{\partial t} = \frac{\partial}{\partial
x}\biggl[V'(x) P(x,t)\biggr] + \frac{\partial^2P(x,t)}{\partial
x^2}.
\end{equation}
Its normalized stationary solution is given by $P^*(x) =
\frac{1}{Z}e^{-V(x)}$, where $Z = \int e^{-V}dx$. We wish to solve
the general initial value problem defined by this equation together
with an initial condition $P_0(x) \equiv P(x,0)$. As discussed in
\sect \ref{sec:Scaling}, by considering deviations from the
equilibrium distribution, we may, without loss of generality,
restrict our discussion to initial conditions with zero
normalization. We therefore assume from now on that
\begin{equation}\label{eq:ExactInitialNormalization}
\int P_0(x)dx = 0.
\end{equation}

To solve the initial value problem defined by such initial
conditions we transform \eqn (\ref{eq:ExactFokkerPlanck}) into an
imaginary-time Schr{\"o}dinger equation via the transformation
\cite{Risken}
\begin{equation}\label{eq:ExactWavefunctionDef}
P(x,t) = e^{-V(x)/2}\psi(x,t).
\end{equation}
The resulting equation for the ``wavefunction'' $\psi$ is
\begin{equation}\label{eq:ExactSchrodinger}
\frac{\partial \psi(x,t)}{\partial t} =
\frac{\partial^2\psi(x,t)}{\partial x^2} - V_s(x) \psi(x,t)
\end{equation}
with the Schr{\"o}dinger potential
\begin{equation}\label{eq:ExactSchrodingerPotentialDef}
V_s(x) \equiv \frac{(V'(x))^2}{4} - \frac{V''(x)}{2}.
\end{equation}
For the potential (\ref{eq:ExactLogPotential}) this gives
\begin{equation}\label{eq:ExactSchrodingerPotential}
V_s(x) = \left\{
\begin{array}{ll} \tilde{V}_s(x) & \mbox{for } |x|<1 \\
\gamma/x^2  & \mbox{for } |x|>1
\end{array} \right.
\end{equation}
with the constant $\gamma = \frac{b}{2}(\frac{b}{2}+1)$. For large
$x$ this equation describes a quantum particle moving in a repulsive
inverse square potential.

\subsection{Eigenfunction representation of the solution}

By separation of variables $\psi(x,t) = \psi_k(x)T(t)$ we find
\begin{equation}
T_k(t) = e^{-k^2t}, \quad k\geq 0
\end{equation}
which yields the time-independent Schr{\"o}dinger equation
\begin{equation}\label{eq:ExactTimeIndSchrodinger}
-V_s(x) \psi_k(x) + \psi_k''(x) = -k^2 \psi_k(x).
\end{equation}

The zero energy (i.e., $k=0$) eigenfunction, which corresponds to
the steady-state solution of the Fokker-Planck equation, is
\begin{equation}\label{eq:ExactZeroWavefunction}
\psi^*(x) \equiv \psi_{k=0}(x) = {\textstyle\frac{1}{\sqrt{Z}}}
e^{-V(x)/2}
\end{equation}
where the normalization $Z^{-1/2}$ ensures that $\int\psi^*(x)^2dx =
1$. The rest of the eigenfunctions can be chosen to be either even
or odd, since $V_s(x)$ is a symmetric potential. Denote the even
eigenfunctions by $\psi_{+,k}(x)$ and the odd by $\psi_{-,k}(x)$,
with $k>0$. These eigenfunctions are
\begin{eqnarray}\label{eq:ExactWavefunctions}
\psi_{\pm,k}(x) = c_\pm(k)\, \left\{
\begin{array}{ll} \phantom{\pm}\tilde{\psi}_{\pm,k}(x), &  |x|<1 \\
\phantom{\pm}\sqrt{|x|}\bigl[c_{\pm,J}(k) J_\rho(k|x|) +
c_{\pm,Y}(k)
Y_\rho(k|x|)\bigr], & \phantom{|}x\phantom{|}>1 \\
\pm \sqrt{|x|}\bigl[c_{\pm,J}(k) J_\rho(k|x|) + c_{\pm,Y}(k)
Y_\rho(k|x|)\bigr], & \phantom{|}x\phantom{|}<-1
\end{array} \right.  \nonumber \\
\end{eqnarray}
where $J_\rho$ and $Y_\rho$ are Bessel functions of the first and
second kind of order $\rho = (b+1)/2$, and $\tilde{\psi}_{\pm,k}(x)$
are the even and odd eigenfunctions of the potential
$\tilde{V}_s(x)$. We choose to normalize $\tilde{\psi}_{\pm,k}$ by
demanding $\tilde{\psi}_{\pm,k}(1) = 1$. The constants $c_{\pm,J}$
and $c_{\pm,Y}$ can be found by proper continuity requirements on
the eigenfunctions at $x=1$. Continuity of the probability $P(x,t)$
and of the probability current $J = V' P + \frac{\partial
P}{\partial x} = (\frac{V'}{2}\psi + \frac{\partial\psi}{\partial
x})e^{-V/2}$ (see \eqn (\ref{eq:ExactFokkerPlanck})) dictate that
for small $\epsilon$
\begin{eqnarray}
\psi_{\pm,k}(1+\epsilon) = \psi_{\pm,k}(1-\epsilon) + O(\epsilon) \nonumber \\
\psi_{\pm,k}'(1+\epsilon) + \frac{b}{2} = \psi_{\pm,k}'(1-\epsilon)
+ \frac{\tilde{V}'(1)}{2} + O(\epsilon).
\end{eqnarray}
This in turn gives for $c_{\pm,J}$ and $c_{\pm,Y}$
\begin{eqnarray}\label{eq:ExactC12}
c_{\pm,J}(k) &= \frac{\pi}{2}\Bigl(\bigl[{\textstyle
\frac{1+b-\tilde{V}'(1)}{2}}-v_\pm(k)\bigr]Y_\rho(k)+kY'_\rho(k)\Bigr),
\nonumber \\
c_{\pm,Y}(k) &=
\frac{\pi}{2}\Bigl(\bigl[v_\pm(k)-{\textstyle\frac{1+b-\tilde{V}'(1)}{2}}\bigr]
J_\rho(k)-kJ'_\rho(k)\Bigr),
\end{eqnarray}
where we have defined $v_\pm(k) \equiv \tilde{\psi}'_{\pm,k}(1) =
v_{\pm,0} + v_{\pm,2} k^2 + \ldots$ (below we show that this series
indeed contains only even powers of $k$). The overall normalization
$c_\pm(k)$, chosen so that for large $x$ the eigenfunctions satisfy
$\psi_{\pm,k}(x) \sim \pi^{-1/2}\sin(kx - \phi_{\pm,k})$ with some
phases $\phi_{\pm,k}$, is
\begin{equation}\label{eq:ExactCk}
c_\pm(k) = \biggl(\frac{k/2}{c_{\pm,J}(k)^2 +
c_{\pm,Y}(k)^2}\biggr)^{1/2}.
\end{equation}
This choice of normalization guarantees the completeness relation
\begin{equation}
\psi^*(x)\psi^*(x') + \int_{0}^\infty
\Bigl[\psi_{+,k}(x)\psi_{+,k}(x')+ \psi_{-,k}(x)\psi_{-,k}(x')\Bigr]
dk = \delta(x-x').
\end{equation}

Using these eigenfunctions and the definition
(\ref{eq:ExactWavefunctionDef}), we can write down the solution to
the original Fokker-Planck equation (\ref{eq:ExactFokkerPlanck}) for
any initial condition $P_0(x)$. Denoting this solution by
$P(x,t|P_0)$, we have
\begin{equation}\label{eq:ExactInergralSolution}
\fl \qquad P(x,t|P_0) = \sum_{\pm} e^{-V(x)/2}\int_0^{\infty} dk \,
\alpha_{\pm}(k) \psi_{\pm,k}(x) e^{-k^2 t}  +
\frac{e^{-V(x)}}{{Z}}\int_{-\infty}^\infty P_0(x_0)dx_0
\end{equation}
where the amplitudes $\alpha_{\pm}(k)$ are given by the projection
of the initial condition on the appropriate eigenfunctions
\begin{equation}\label{eq:ExactGofK}
\alpha_{\pm}(k) \equiv \int_{-\infty}^\infty dx_0\,P_0(x_0)
e^{V(x_0)/2}\psi_{\pm,k}(x_0).
\end{equation}
The second term in the rhs of \eqn (\ref{eq:ExactInergralSolution})
is obtained by projecting $P_0(x)$ on $\psi^*(x)$, i.e.,
substituting (\ref{eq:ExactZeroWavefunction}) into the expression
$e^{-V(x)/2}\psi^*(x)\int dx_0\, P_0(x_0) e^{V(x_0)/2}\psi^*(x_0)$.
For the zero-normalization initial condition
(\ref{eq:ExactInitialNormalization}) which we consider, this term
vanishes.

\subsection{Eigenfunctions and amplitudes at small $k$}

We are interested in the long-time behavior of the solution. The
$e^{-k^2 t}$ term in the first integral of
(\ref{eq:ExactInergralSolution}) implies that when $t\gg 1$ only
small values of $k$ will contribute to the integral. We are
therefore led to investigate the small $k$ behavior of the
amplitudes $\alpha_\pm(k)$ (which according to the definition
(\ref{eq:ExactGofK}) may depend on the initial condition).

First, let us examine the small $k$ asymptotics of the constants
$c_{\pm,J}(k)$, $c_{\pm,Y}(k)$ and $c_\pm(k)$. For small $k$, the
eigenfunctions in the region $-1<x<1$ can be expanded as a power
series
\begin{equation}\label{eq:ExactPsiTildeSeries}
\tilde{\psi}_{\pm,k}(x) = \tilde{\psi}_{\pm,0}(x) + k^2 h_{\pm,2}(x)
+ k^4 h_{\pm,4}(x) + \ldots.
\end{equation}
This expansion is uniform in $x$ in this region, and includes only
even powers of $k$ as the eigenvalue problem
(\ref{eq:ExactTimeIndSchrodinger}) is even in $k$. The zeroth order
terms are
\begin{eqnarray}\label{eq:ExactPsiTildeZero}
\tilde{\psi}_{+,0}(x) &= \frac{\psi^*(x)}{\psi^*(1)} \nonumber \\
\tilde{\psi}_{-,0}(x) &= \frac{\psi^*(x)\int_0^x
e^{V(y)}dy}{\psi^*(1)\int_0^1 e^{V(y)}dy}.
\end{eqnarray}
Here, $\psi^*(x)$ is given by (\ref{eq:ExactZeroWavefunction}), from
which we can deduce that $v_{+,0} = \frac{{\psi^*}'(1)}{\psi^*(1)} =
-\frac{\tilde{V}'(1)}{2}$, and similarly $v_{-,0} =
-\frac{\tilde{V}'(1)}{2} + \frac{e^{V(1)}}{\int_0^1 e^{V(y)}dy}$. By
substituting the expansion (\ref{eq:ExactPsiTildeSeries}) in the
Schr{\"o}dinger equation (\ref{eq:ExactSchrodinger}) and continuing
the perturbative calculation to the next order, it can also be shown
that $v_{+,2} = h_{+,2}'(1) = -\int_{-1}^1 \frac{e^{-V}}{2} dx$,
which, together with $\int_{1}^\infty e^{-V}dx = 1/(b-1)$, yields $Z
\equiv \int_{-\infty}^\infty e^{-V}dx = 2/(b-1)-2v_{+,2}$.
Substituting these in (\ref{eq:ExactC12}) and (\ref{eq:ExactCk}) we
find
\begin{eqnarray}\label{eq:ExactCSeries}
c_{+,J}(k) = - Z
\Gamma(\rho)\biggl(\frac{2}{k}\biggr)^{\rho-2}\bigl[1 +
O(k^2)\bigr],
\nonumber \\
c_{+,Y}(k) = \frac{-\pi}{\Gamma(\rho)}
\biggl(\frac{k}{2}\biggr)^{\rho} \bigl[1 + O(k^2)\bigr] \\
c_+(k) =
\frac{-1}{Z\Gamma(\rho)}\biggl(\frac{k}{2}\biggr)^{\rho-\frac{3}{2}}
\bigl[1 + \bigOtwo{k}{2}{b-1}\bigr], \nonumber
\end{eqnarray}
and similarly $c_{-,J}(k) \sim k^{-\rho}$, $c_{-,Y}(k) \sim k^\rho$
and $c_{-}(k) \sim k^{\rho+1/2}$ (the coefficients of the latter
three are omitted because they will not be used below). With these,
together with the known asymptotics of the Bessel functions
\cite{AbramowitzStegun}, we may rewrite the eigenfunctions
(\ref{eq:ExactWavefunctions}) for $x>1$ and $k\ll 1$ as
\begin{eqnarray}\label{eq:ExactWavefunctionsSmallK}
\fl \qquad \psi_{\pm,k}(x>1) &\approx \sqrt{\frac{kx}{2}}J_\rho(kx)
+ \frac{\pi}{Z\Gamma^2(\rho)}\biggl(\frac{k}{2}\biggr)^{b-1}
\sqrt{\frac{kx}{2}}Y_\rho(kx) = \nonumber \\
&= \frac{(kx)^{\frac{b}{2}+1}}{\Gamma(\rho+1)2^{\frac{b}{2}+1}}
\Bigl[1+O\bigl((kx)^2\bigr)\Bigr] - \frac{k^{\frac{b}{2}-1}
x^{-\frac{b}{2}}}{Z\Gamma(\rho)2^{\frac{b}{2}-1}}
\Bigl[1+O\bigl((kx)^2\bigr)\Bigr].
\end{eqnarray}

In order to study the amplitudes (\ref{eq:ExactGofK}), we must make
some assumptions about the initial condition $P_0(x)$. Below we
assume that $P_0(x)$ is asymptotically symmetric for large $|x|$,
i.e.,
\begin{equation}\label{eq:ExactInitial}
P_0(x\gg 1) = P_0(-x\ll -1) \sim P^*(x)\cdot A|x|^{-a}.
\end{equation}
This assumption is made solely for notational simplicity. In
general, one could have $P_0(x\to \pm \infty) \sim P^*(x)\cdot
A_\pm|x|^{-a_\pm}$. The calculation which we present below can be
repeated for this more general case, resulting in different scaling
behaviors for positive and negative $x$'s, in which case only the
smaller of $a_+$ and $a_-$ dominates the eventual long-time
behavior. We further assume, without loss of generality, that for
all $|x|>1$, not just for large $x$, the initial condition is
already close to its asymptotic form, i.e., $P_0(|x|> 1) \approx
P^*(x)\cdot A|x|^{-a}$ (one can rescale $x$ and $t$ to ensure that
this is the case; note that such a rescaling entails a redefinition
of $\tilde{V}$, $\tilde{\psi}$ and $v_{\pm}(k)$).

When the initial conditions are asymptotically symmetric, the small
$k$ behavior of  $\alpha_\pm(k)$ is determined as follows.
Separating the integration in (\ref{eq:ExactGofK}) to three
integrals and substituting Equations (\ref{eq:ExactLogPotential})
and (\ref{eq:ExactWavefunctions}), we can write
\begin{eqnarray}\label{eq:ExactGofKIntegral}
\alpha_+(k) &= c_+(k) \Bigl[2 I_1(k) + 2 I_2(k) + I_{+,3}(k)\Bigr] \nonumber \\
\alpha_-(k) &= c_-(k) I_{-,3}(k),
\end{eqnarray}
with
\begin{eqnarray}
I_1(k) = \int_{1}^\infty dx\,c_{+,J}(k)J_\rho(k x) x^\rho P_0(x)
\nonumber
\\
I_2(k) = \int_{1}^\infty dx\,c_{+,Y}(k)Y_\rho(k x) x^\rho P_0(x)
\\
I_{\pm,3}(k) = \int_{-1}^{1} dx\, \tilde{\psi}_{\pm,k}(x)
e^{\tilde{V}(x)/2} P_0(x) \nonumber
\end{eqnarray}
(note that $P_0(x)$ need not be symmetric for $-1<x<1$). Changing
the integration variable in the first integral to $z = k x$ and
substituting Equations (\ref{eq:ExactCSeries}),
(\ref{eq:ExactInitial}) and $P^*(x) = \frac{1}{Z}e^{-V(x)}$, yields
\begin{equation}\label{eq:ExactI1a}
I_1(k) \approx -A \Gamma(\rho)2^{\rho-2}\cdot k^a \cdot
\int_{k}^\infty dz\,z^{(1-b-2a)/2}J_\rho(z).
\end{equation}
The latter integral converges to a finite value when $k\to 0$ if
$(1-b-2a)/2+\rho
> -1$, or equivalently if $a<2$; otherwise it diverges with $k$. Its
asymptotic behavior is given, to leading order in $k$, by
\begin{equation}\label{eq:ExactI1b}
\int_{k}^\infty dz\,z^{(1-b-2a)/2}J_\rho(z) \sim \left\{
\begin{array}{ll}
\frac{2^{1-\rho-a}\Gamma(1-\frac{a}{2})}{\Gamma(\frac{a+b+1}{2})}\phantom{\biggl)}
& \mbox{when } a<2 \\
-\frac{2^{-(b+1)/2}}{\Gamma(\frac{b+3}{2})}\log k\phantom{\biggl)} & \mbox{when } a = 2\\
\frac{1}{(a-2)2^{\rho}\Gamma(\rho+1)}\,k^{2-a}\phantom{\biggl)}  &
\mbox{when } a>2
\end{array} \right..
\end{equation}
To evaluate $I_2(k)$, for all $k x \ll 1 $ we can approximate
$c_{+,Y}(k) Y_\rho(k x) = x
^{-\rho}\bigl[1+O\bigl((kx)^2\bigr)\bigr]$ (using \eqn
(\ref{eq:ExactCSeries}) and the known asymptotics of the Bessel
function). Fixing an $\epsilon \ll 1$, the integral is evaluated as
\begin{eqnarray}
I_2(k) &= \int_1^{\epsilon/k}P_0(x) dx + \int_{\epsilon/k}^\infty
c_{+,Y}(k) Y_\rho(kx) x^\rho P_0(x) dx +
O(k^2)= \nonumber \\
&= \int_1^{\infty}P_0(x) dx + \int_{\epsilon/k}^\infty [c_{+,Y}(k)
Y_\rho(kx) x^\rho - 1]P_0(x) dx + O(k^2).
\end{eqnarray}
Changing once again the integration variable to $z = k x$ reveals
that
\begin{equation}\label{eq:ExactI2}
I_2(k) = \int_1^{\infty}P_0(x)dx + O(k^2, k^{a+b-1}).
\end{equation}
The integrals $I_{\pm,3}(k)$ are evaluated using the expansion
(\ref{eq:ExactPsiTildeSeries})--(\ref{eq:ExactPsiTildeZero}):
\begin{eqnarray}\label{eq:ExactI3}
I_{+,3} &= \int_{-1}^1 P_0(x) dx + O(k^2) \nonumber \\
I_{-,3} &= \int_{-1}^1 \frac{\int_0^x e^{V(y)}dy }{\int_0^1
e^{V(y)}dy}P_0(x)dx + O(k^2).
\end{eqnarray}
Combining (\ref{eq:ExactI1a})--(\ref{eq:ExactI1b}),
(\ref{eq:ExactI2}) and (\ref{eq:ExactI3}) into Equations
(\ref{eq:ExactCSeries}) and (\ref{eq:ExactGofKIntegral}), and
remembering that $\int_{-\infty}^{\infty} P_0(x)dx = 0$, finally
yields, to leading order,
\begin{equation}\label{eq:ExactGofKFinal}
\alpha_{\pm}(k) \sim C_\pm k^{\nu_\pm} \qquad \mbox{when } a \neq 2,
\end{equation}
with
\begin{equation}\label{eq:ExactGofKExponent}
\nu_{-} = \frac{b}{2} + 1, \quad \mbox{and} \quad \nu_+ = \left\{
\begin{array}{ll}
\frac{b}{2} + 1 - (2-a)\;{}
& \mbox{if } a<2 \\
\frac{b}{2} + 1 & \mbox{if } a>2
\end{array} \right..
\end{equation}
The constants $C_\pm$ are non-universal (i.e., they depend on the
full forms of the potential $V(x)$ and of the initial condition
$P_0(x)$), except when $a<2$, where
\begin{equation}\label{eq:ExactGofKConstant}
C_{+}(a<2) = \frac{A}{Z}\cdot
\frac{2^{1-a-b/2}\Gamma(1-\frac{a}{2})}{\Gamma(\frac{b+a+1}{2})}.
\end{equation}
When $a=2$, $\alpha_-(k)$ is still given by \eqns
(\ref{eq:ExactGofKFinal})--(\ref{eq:ExactGofKExponent}), but the
expression for $\alpha_{+}(k)$ is replaced with
\begin{equation}\label{eq:ExactGofKLog}
\alpha_{+}(k) \sim - \frac{A}{Z}\cdot
\frac{2^{-b/2}}{\Gamma(\frac{b+3}{2})} \cdot k^{\frac{b}{2} + 1}\log
k \qquad \mbox{when } a=2.
\end{equation}

\subsection{Late-time scaling solutions}

Once the asymptotic forms of $\psi_{\pm,k}(x)$ and $\alpha_\pm(k)$
for small $k$ (\eqns (\ref{eq:ExactPsiTildeZero}),
(\ref{eq:ExactWavefunctionsSmallK}),
(\ref{eq:ExactGofKFinal})--(\ref{eq:ExactGofKLog})) are known, they
can be substituted into equation (\ref{eq:ExactInergralSolution}).
When $a\neq 2$, changing the integration variable to $q=k t^{1/2}$
yields in the region $|x|<1$
\begin{eqnarray}\label{eq:ExactCutoffXIntegral}
\fl \int_0^{\infty} dk \, \alpha_{+}(k) \psi_{+,k}(x) e^{-k^2 t} =
t^{-1/2}\int_0^{\infty} dq \, \alpha_{+}(qt^{-1/2})
c_+(qt^{-1/2}) \tilde{\psi}_{+,qt^{-1/2}}(x) e^{-q^2} = \nonumber \\
= - \frac{C_+
t^{-\frac{b+2\nu_\pm}{4}}}{Z\Gamma(\frac{b+1}{2})}\frac{\psi^*(x)}{\psi^*(1)}
\int_0^\infty q^{\frac{b}{2}+\nu_+-1} e^{-q^2} dq
\Bigl(1+O(t^{-1})\Bigr).
\end{eqnarray}
A similar calculation for the odd eigenfunctions shows that their
contribution is negligible in comparison with
(\ref{eq:ExactCutoffXIntegral}) for all $|x|<1$. In the region
$|x|\geq 1$, we similarly have
\begin{eqnarray}\label{eq:ExactSmallXIntegral}
\fl \int_0^{\infty} dk \, \alpha_{\pm}(k) \psi_{\pm,k}(x) e^{-k^2 t}
= \frac{C_\pm t^{-\frac{b+2\nu_\pm}{4}}x^{-\frac{b}{2}}}
{\Gamma(\frac{b+3}{2})2^{\frac{b}{2}+1}}  \nonumber \\
\times \biggl[ \frac{x^{b+1}}{t} \int_0^\infty
q^{\frac{b}{2}+\nu_\pm+1} e^{-q^2} dq - \frac{2(b+1)}{Z}
\int_0^\infty q^{\frac{b}{2}+\nu_\pm-1} e^{-q^2} dq
\biggr]\Bigl(1+O(x^2 t^{-1})\Bigr).\nonumber \\
\end{eqnarray}
When $a = 2$, equations (\ref{eq:ExactCutoffXIntegral}) and
(\ref{eq:ExactSmallXIntegral}) have a similar form but are
multiplied by an overall $\log t$ correction factor.

As long as $|x|\ll t^{1/2}$, the higher order terms in \eqns
(\ref{eq:ExactCutoffXIntegral}) and (\ref{eq:ExactSmallXIntegral})
can be dropped. Using the identity $\int_0^\infty q^\mu e^{-q^2}dq =
\Gamma(\frac{\mu + 1}{2})/2$ then leads to
\begin{equation}\label{eq:ExactFinalScalingSmallX}
\fl \qquad \frac{P(x\ll \sqrt{t},t|P_0)}{P^*(x)} \approx C\,\left\{
\begin{array}{ll}
t^{-\frac{b+a-1}{2}}
\Bigl[-\frac{4(b+1)}{Z(b+a-1)}+\frac{|x|^{b+1}}{t}\Bigr]\phantom{\bigg)}
& \mbox{when } a<2 \\
t^{-\frac{b+1}{2}}\log t \,
\Bigl[-\frac{4}{Z}+\frac{|x|^{b+1}}{t}\Bigr]\phantom{\bigg)}
& \mbox{when } a = 2\\
t^{-\frac{b+1}{2}}
\Bigl[-\frac{4}{Z}+\frac{|x|^{b+1}}{t}\Bigr]\phantom{\bigg)} &
\mbox{when } a>2
\end{array} \right.,
\end{equation}
where
\begin{equation}\label{eq:ExactFinalConstant}
C = \left\{
\begin{array}{ll}
\frac{A\Gamma(1-\frac{a}{2})}
{2^{b+a+1}\Gamma(\frac{b+3}{2})}\phantom{\bigg)}
& \mbox{when } a<2 \\
\frac{A} {2^{b+3}\Gamma(\frac{b+3}{2})}\phantom{\bigg)} & \mbox{when
} a = 2
\end{array} \right.,
\end{equation}
and $C$ is non-universal when $a>2$ (compare with \eqns
(\ref{eq:FinalScalingAnsatz})--(\ref{eq:FinalSmallXScalingFunction})
and (\ref{eq:FinalBeta})--(\ref{eq:FinalConstant})). Note that \eqn
(\ref{eq:ExactFinalScalingSmallX}) holds even in the region $|x|<1$,
where the potential is not logarithmic.

When $x = O(t^{1/2})$, the higher order terms in \eqn
(\ref{eq:ExactSmallXIntegral}) cannot be neglected. They are taken
into account by leaving the Bessel functions in \eqn
(\ref{eq:ExactWavefunctionsSmallK}) unexpanded when substituting in
\eqn (\ref{eq:ExactInergralSolution}). Changing the integration
variable once again to $q=k t^{1/2}$, and substituting $u \equiv
xt^{-1/2}$, yields when $a\neq 2$
\begin{eqnarray}
\fl \int_0^{\infty} dk \, \psi_{\pm,k}(x) \alpha_{\pm}(k) e^{-k^2 t} \approx
\nonumber \\
\approx \frac{C_\pm
u^{1/2}t^{-\frac{\nu_\pm+1}{2}}}{\sqrt{2}}\int_0^\infty
q^{\frac{1}{2}+\nu_\pm} \Bigl[J_\rho(qu)+\frac{\pi
q^{b-1}}{Z\Gamma^2(\rho)2^{b-1}}
Y_\rho(qu)t^{-\frac{b-1}{2}}\Bigr]e^{-q^2}dq, \nonumber \\
\end{eqnarray}
and a similar expression with a $\log t$ correction when $a=2$. The
second term in the square brackets is negligible at late times.
Using the identity \cite{AbramowitzStegun}
\begin{eqnarray}
\int_0^\infty q^\mu J_\rho(qu) e^{-q^2} dq = { \frac{2^{-(\rho+1)}
\Gamma(\frac{1+\rho+\mu}{2})}{\Gamma(1+\rho)}}\, u^\rho \,_1\!F_1
\Bigl(\frac{1+\rho+\mu}{2};1+\rho;-\frac{u^2}{4}\Bigr) \nonumber \\
\end{eqnarray}
together with (\ref{eq:HypergeomExpIdentity}), we finally arrive at
the scaling solution
\begin{equation}\label{eq:ExactFinalScaling}
\fl \qquad \frac{P(ut^{1/2},t|P_0)}{P^*(ut^{1/2})} \approx C\,
\left\{
\begin{array}{ll}
u^{b+1}
\,_1\!F_1\left(\frac{b+a+1}{2};\frac{b+3}{2};-\frac{u^2}{4}\right)\cdot
t^{-\frac{a}{2}}\phantom{\biggl)}
& \mbox{when } a<2 \\
u^{b+1} e^{-\frac{u^2}{4}} \cdot t^{-1}\log
t \phantom{\biggl)} & \mbox{when } a = 2\\
u^{b+1} e^{-\frac{u^2}{4}} \cdot t^{-1} \phantom{\biggl)}
  & \mbox{when } a>2
\end{array} \right.,
\end{equation}
where the constant $C$ is the same as in
(\ref{eq:ExactFinalConstant}) (compare with \eqns
(\ref{eq:FinalScalingAnsatz})--(\ref{eq:FinalScalingFunction}) and
(\ref{eq:FinalBeta})--(\ref{eq:FinalConstant})).

We have repeated the calculation of this appendix also for the case
considered in \sect \ref{sec:NonConservingBC} of absorbing boundary
conditions at the origin. This lengthy but straightforward
calculation, which we do not present here, recovers \eqns
(\ref{eq:FinalSmallXAnsatzAbsorbing})--(\ref{eq:FinalScalingAnsatzAbsorbing})
and yields the logarithmic corrections when $a=2$. In particular,
the calculation reveals that \eqn (\ref{eq:ExactFinalScaling}) holds
regardless of the boundary condition.

\section{Calculation of the sum in \eqn
(\ref{eq:ZRPCurrentCorrelator})}\label{sec:ZRPappendix}

In this appendix, we calculate the current correlation function
discussed in \sect \ref{sec:ZRP} using the scaling solution
(\ref{eq:FinalScalingAnsatz}). We begin by splitting the sum in \eqn
(\ref{eq:ZRPCurrentCorrelator}) into three terms,
\begin{equation}\label{eq:ZRPCurrentCorrelator2}
C(t) = S_1(t) + S_2(t)+S_3(t),
\end{equation}
where we define
\begin{eqnarray}\label{eq:ZRPSsDef}
S_1(t) &\equiv \sum_{m=n_1(t)+1}^\infty
\Bigl[P(m,t|P_0)-P^*(m)\Bigr]\bigl(w(m)-1\bigr) \nonumber \\
S_2(t)&\equiv \sum_{\stackrel{\hphantom{\scriptstyle
m=n_1(t)+1}}{m=1}}^{n_1(t)}
\Bigl[P(m,t|P_0)-P^*(m)\Bigr]w(m) \\
S_3(t) &\equiv \sum_{m=n_1(t)+1}^\infty
\Bigl[P(m,t|P_0)-P^*(m)\Bigr] = - \sum_{m=0}^{n_1(t)}
\Bigl[P(m,t|P_0)-P^*(m)\Bigr]. \nonumber
\end{eqnarray}
Here, $n_1(t)$ is chosen to satisfy $t^{1/(b+1)} \ll n_1(t) \ll
t^{1/2}$, and in the last equality we have used the normalization
condition $\sum_m P^*(m) = \sum_m P(m,t|P_0)=1$.

As discussed in \sect \ref{sec:ZRP}, at large times, the terms in
the square brackets in (\ref{eq:ZRPSsDef}) can be replaced by the
scaling solution (\ref{eq:FinalScalingAnsatz}) and
(\ref{eq:FinalBeta}) with $a=1$. In the first sum, the square
brackets are replaced with the large-$x$ scaling function, yielding
\begin{eqnarray}
S_1(t) &=
\sum_{u=\frac{n_1(t)+1}{\sqrt{t}},\frac{n_1(t)+2}{\sqrt{t}},\ldots}^\infty
\Bigl[P^*(u\sqrt{t})Ct^{-1/2}f_{1/2}(u)\Bigr]\bigl(w(u\sqrt{t})-1\bigr)
\approx \nonumber\\
&\approx \frac{Cb}{\mathcal{Z}}\,t^{-\frac{b+1}{2}}\int_0^\infty
\,_1\!F_1\left(\frac{1+b+a}{2};\frac{b+3}{2};-\frac{u^2}{4}\right)
du = \frac{\pi \Gamma(\frac{1+b}{2})} {\mathcal{Z}
2^{b}\Gamma^2\bigl(\frac{b}{2}\bigr)}\, t^{-\frac{b+1}{2}},
\nonumber \\
\end{eqnarray}
where we have substituted the asymptotic form of $P^*(n)$
(\ref{eq:ZRPStationaryDist}) and the value of the constant $C$ which
is given in (\ref{eq:FinalConstant}).

A similar calculation  is carried out for $S_2$ and $S_3$, this time
using the small-$x$ scaling function. We now show that although
$S_2,S_3 \sim t^{-b/2}$, the two sums cancel each other to leading
order in $t$. To see this, substitute (\ref{eq:FinalScalingAnsatz})
in (\ref{eq:ZRPSsDef}), and use (\ref{eq:ZRPStationaryDist}) to
deduce that $P^*(n)w(n) = P^*(n-1)$. Combining these gives
\begin{eqnarray}\label{eq:ZRPLeadingS2S3}
S_2(t) &\approx \sum_{m=1}^{n_1(t)} t^{-b/2}P^*(m)w(m)\Bigl[C_3 +
C\frac{m^{b+1}}{t}\Bigr] = \nonumber \\
&= \sum_{m=0}^{n_1(t)-1} t^{-b/2}P^*(m)\Bigl[C_3 +
C\frac{(m+1)^{b+1}}{t}\Bigr]
\\
S_3(t) &\approx -\sum_{m=0}^{n_1(t)} t^{-b/2}P^*(m)\Bigl[C_3 +
C\frac{m^{b+1}}{t}\Bigr], \nonumber
\end{eqnarray}
where $C_3$ is given in (\ref{eq:SmallXConstant1}). Therefore,
\begin{eqnarray}
S_2(t)+S_3(t) \approx &\sum_{m=0}^{n_1-1} C
t^{-\frac{b}{2}-1}P^*(m)[(m+1)^{b+1}-m^{b+1}] - {} \nonumber \\
& {} - t^{-\frac{b}{2}}P^*\bigl(n_1(t)\bigr)\Bigl[C_3 +
C\frac{n_1(t)^{b+1}}{t}\Bigr],
\end{eqnarray}
from which it can be shown that, to this order, $|S_2(t)+S_3(t)| \ll
t^{-\frac{b+1}{2}}$. Using a perturbative expansion similar to
(\ref{eq:ExactPsiTildeSeries}), it can be shown that the
contribution from higher order corrections to the scaling form are
also negligible.

Adding the three contributions together, we find that the asymptotic
decay of the correlation function is
\begin{equation}
C(t) \approx S_1(t) \approx \frac{\pi \Gamma(\frac{1+b}{2})}
{\mathcal{Z} 2^{b}\Gamma^2\bigl(\frac{b}{2}\bigr)}\,
t^{-\frac{b+1}{2}}.
\end{equation}


\bibliographystyle{unsrt}
\bibliography{log_diff_long}

\end{document}